\documentclass[preprint2]{aastex}\usepackage{graphicx}
\usepackage{float}
\usepackage[caption = false]{subfig}

\begin{document}

\title{Photometric Trends in the Visible Solar Continuum and \\
    Their Sensitivity to the Center-to-Limb Profile}

\author{C.L. Peck\altaffilmark{1, 3} and M.P. Rast\altaffilmark{2, 3}}
\affil{Department of Physics\altaffilmark{1}, Department of Astrophysical and Planetary Sciences\altaffilmark{2}, Laboratory for Atmospheric and Space Physics\altaffilmark{3} University of Colorado, Boulder, CO 80303}

\begin{abstract}
Solar irradiance variations over solar rotational time-scales are largely determined by the passage of magnetic structures across the visible solar disk.  Variations on solar cycle time scales are thought to be similarly due to changes in surface magnetism with activity. Understanding the contribution of magnetic structures to total solar irradiance and solar spectral irradiance requires assessing their contributions as a function of disk position.  Since only relative photometry is possible from the ground, the contrasts of image pixels are measured with respect to a center-to-limb intensity profile. Using nine years of full-disk red and blue continuum images from the Precision Solar Photometric Telescope at the Mauna Loa Solar Observatory (PSPT/MLSO), we examine the sensitivity of continuum contrast measurements to the center-to-limb profile definition. Profiles which differ only by the amount of magnetic activity allowed in the pixels used to determine them yield oppositely signed solar cycle length continuum contrast trends; either agreeing with the result of Preminger et al. (2011) showing negative correlation with solar cycle or disagreeing and showing positive correlation with solar cycle.  Changes in the center-to-limb profile shape over the solar cycle are responsible for the contradictory contrast results, and we demonstrate that the lowest contrast structures, internetwork and network, are most sensitive to these. Thus the strengths of the full-disk, internetwork, and network photometric trends depend critically on the magnetic flux density used in the quiet-sun definition.  We conclude that the contributions of low contrast magnetic structures to variations in the solar continuum output, particularly to long-term variations, are difficult, if not impossible, to determine without the use of radiometric imaging.

\end{abstract}

\keywords{Sun: photosphere; Sun: activity}

\section{Introduction}

Solar irradiance studies to date have focused on disk integrated radiometry and magnetic structure based irradiance reconstructions.  Space-based measurements of the disk-integrated total solar irradiance (TSI) over the past several decades show $\sim0.1\%$ variation on the time scales of solar rotation and the 11-year activity cycle~\citep[e.g.,][and references therein]{frohlich}.  More recent measurements of the disk-integrated solar spectral irradiance (SSI) have suggested that some wavelengths vary in phase with TSI while others do not~\citep[e.g.,][]{harder}.  Ground based full-disk imaging has achieved relative photometric precision (pixel-to-pixel) of up to 0.1\%~\citep[e.g.,][]{rast}, and these images have been used to understand the contribution of individual magnetic structures to the total and spectral irradiance~\citep[e.g.,][]{fontenla}.

The appearance and disappearance of magnetic structures can account for most of the variation in the TSI observed over the course of a solar cycle. Sunspots and pores have negative contrast against the background disk, while smaller scale magnetic elements generally contribute positively, and more so toward the limb, though whether the net contribution including a larger surrounding area is also positive has been recently brought into question by radiative magnetohydrodynamic simulations of small flux elements~\citep{thaler}. Empirical two-component models based on a sunspot deficit (from observed areas and locations) and a facular excess (determined using chromospheric proxies and/or the observed facular area from filter-gram images or longitudinal magnetograms) can account for up to $90$\% of the observed TSI variation~\citep[e.g.,][]{frohlichlean,krivova2006,ball}.  These models do particularly well reproducing irradiance variations on solar rotation time scales, where deep drops in irradiance correspond to the passage of sunspots and enhanced irradiance results from large regions of faculae particularly at the limb. The positive faculae play a dominate role on cycle time scales or longer. The origin or existence of any longer-term secular variation, however, remains unknown. Suggestions for such a long-term component range from the enhancement by change in the number of small-scale elements~\citep{ermolli,ortiz} to underlying thermodynamic changes~\citep{kuhn1998}. 

This paper is concerned with solar spectral irradiance variations, which are more complex. The magnitude and even the sign of the spectral contribution of small scale magnetic structures depends critically on the wavelength being observed, the strength and distribution of the magnetic field within the structure, and the disk position at which the structure is found. Models of the solar spectral irradiance typically proceed semi-empirically, using a multi-component decomposition of solar images into structure classes, one-dimensional model atmospheres of each structure class, and spectral synthesis via a sum over contributions of each structure as a function of its disk position~\citep[e.g.,][]{fon11}.  Solar images are used in such analysis in two ways, for structure identification, usually employing Ca ll K emission or magnetic flux density measurements, and for empirical construction of the underlying one dimensional model atmospheres based on the wavelength dependent center-to-limb variations (CLV). 

As there are no full-disk radiometric images available, photometric images from ground based observations are employed in one or both of these steps.  Unfortunately, analysis of the photometric image data requires a definition of the quiet-sun intensity against which the contrasts of all structures on the disk are measured. This means that structure identification and contrast measurements are made relative to a somewhat arbitrary definition of the quiet-sun CLV.  Temporal variations in that CLV profile can result from error or inconsistencies in the identification of the quiet-sun pixels, changes in the filling factor of unresolved magnetic flux elements, or true changes in the underlying thermodynamic structure of the star.  These are difficult to separate, and any method using contrast images in the synthesis of the solar spectral output faces inherent difficulty in distinguishing actual trends in magnetic structure contributions from temporal variations in the CLV profile.

Work by \citet{ermolli} and \citet{ortiz} investigated the possibility that small scale magnetic structures, particularly magnetic network, introduce long term variability in the TSI.  Using ground-based PSPT observations from the Osservatorio Astronomico di Roma (OAR), \citet{ermolli} found cycle dependent enhancement of  network contributions, while \citet{ortiz} using intensity images from SOHO/MDI, found no such enhancement.  The balloon-borne Solar Bolometric Imager experiment~\citep[SBI,][]{foukal} found no indication of brightness variations in the quiet-sun contributions. While Ortiz notes that the network discrepancy could result from the differing structure definitions, the limited spatial resolution  of both the MDI and SBI instruments (4 and 5 arc seconds, respectively) and the limited photometric precision (about 1\%) of the MDI images may also have contributed.  At low spatial resolution and photometric precision, network and quite-sun contributions are difficult to disentangle and depend sensitively on the CLV used in their identification.

There have also been efforts to use full-disk photometric imagery (contrast images) directly to understand irradiance trends without decomposing the images into magnetic components. \citet{prem1} at the San Fernando Observatory computed disk integrated red and blue continuum contrasts, finding them to be slightly anti-correlated with solar activity. They concluded that the Sun's visible continuum is spot-dominated and therefore diminishes at solar maximum, a result contrary to SSI model reconstructions which suggest that the increased abundance of bright magnetic elements, such as faculae, with increased activity outweighs the sunspot deficit~\citep[e.g.,][]{lean}. ~\citet{prem1} concluded that the reduction in the continuum output with increased magnetic activity is in qualitative agreement with the SIM spectral irradiance measurements which suggest that these wavelengths respond out of phase over the solar cycle~\citep{harder}, however they are in disagreement with SOHO/VIRGO measurements which find the visible continuum to vary in phase~\citep{wehrli}.

These contradictory results, both that of the small magnetic structure contribution and the full-disk continuum trends, motivate our investigation of the role of the CLV in the measurements.  We re-examine the disk integrated continuum contrast as a function of solar cycle using two different definitions of the quiet-sun reference. We find that the trends are highly dependent on the CLV profile employed, and that small changes in the profile shape can remove or reverse the observed trends. We also evaluate the role of the CLV in determining contrast trends of specific magnetic structure classes. We find that the very abundant low contrast features, active network, network, and internetwork, are those most susceptible to variations in the CLV.  This has implications for both relative photometric measurements of their variation and the disk integrated contrast total which they dominate. We examine the effects of dramatically reducing the contribution of weak magnetic structures to the quiet-sun CLV reference, and find that the strengths of the photometric trends depend critically on the magnetic flux density used in the quiet-sun definition, both through the identification scheme and via weak field continuum contributions. 

\begin{figure*}[t]
	\subfloat{\includegraphics[width=85mm,height=30mm, trim=25mm 45mm 0mm 35mm, clip=true]{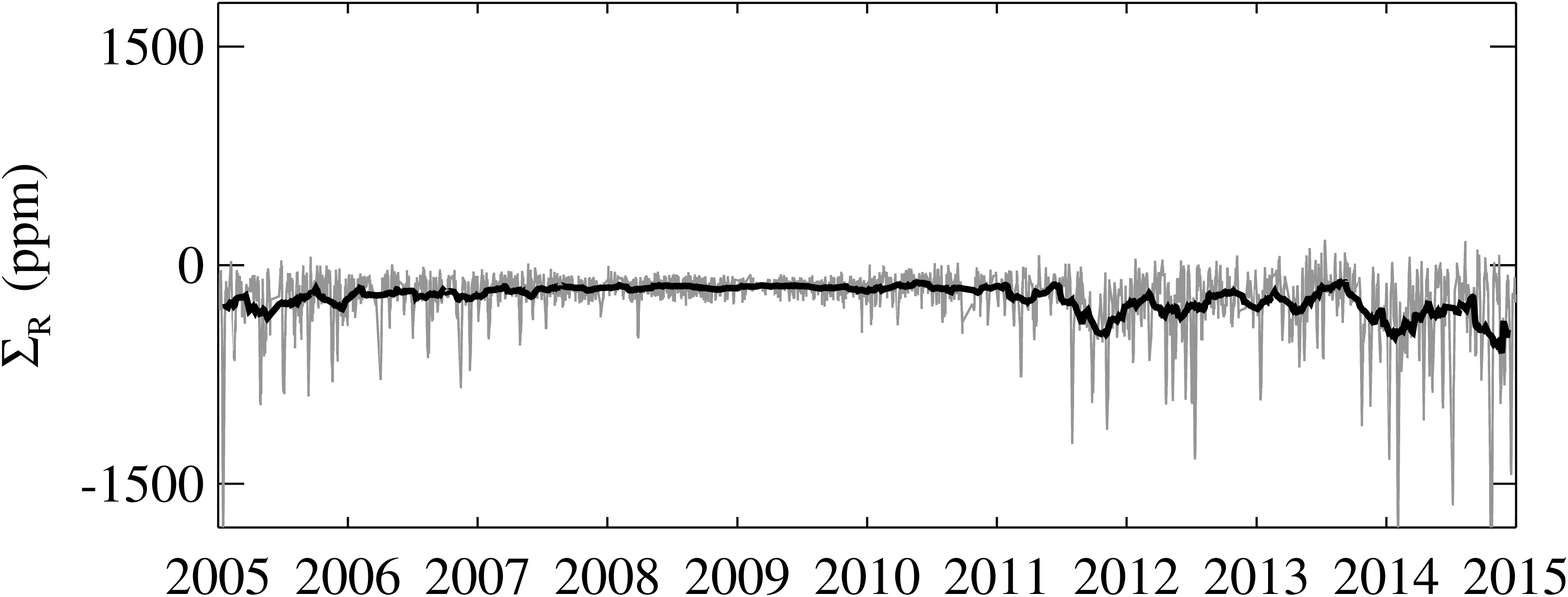}}
	\subfloat{\includegraphics[width=85mm,height=30mm, trim=25mm 45mm 0mm 35mm, clip=true]{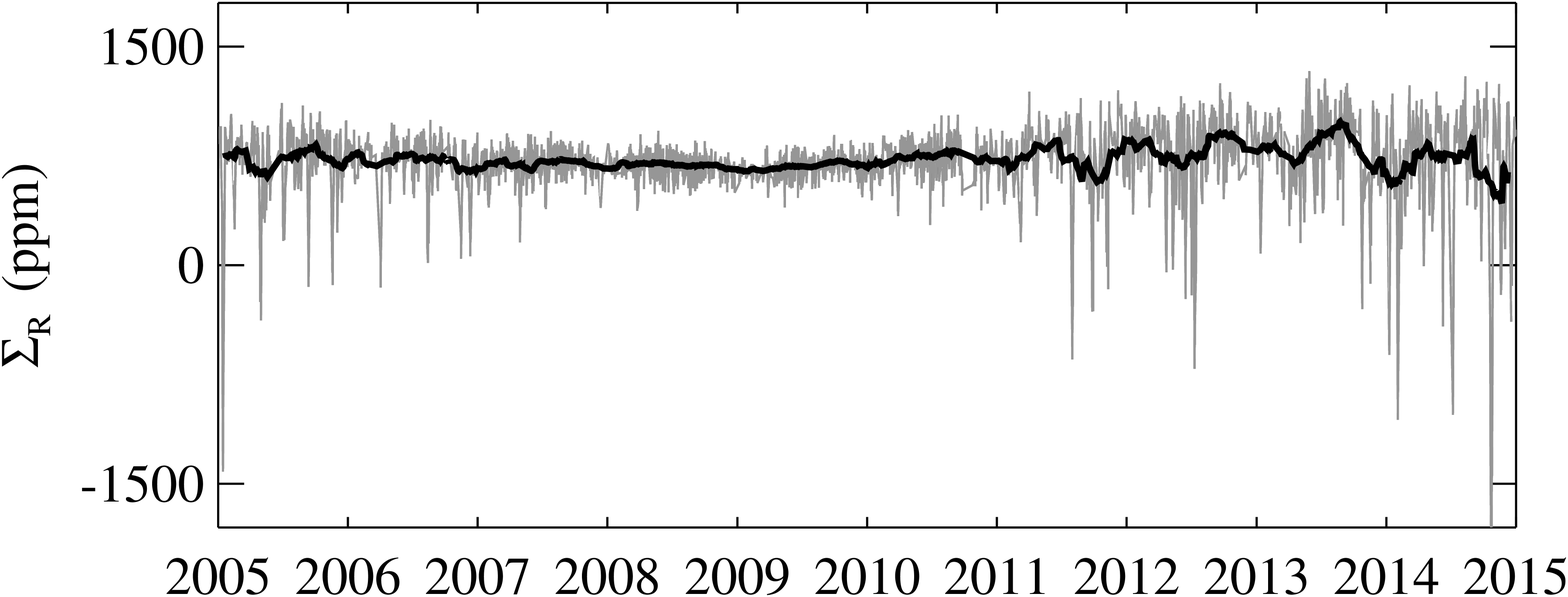}}\\
	\subfloat{\includegraphics[width=85mm,height=32mm, trim=25mm 22mm 0mm 35mm, clip=true]{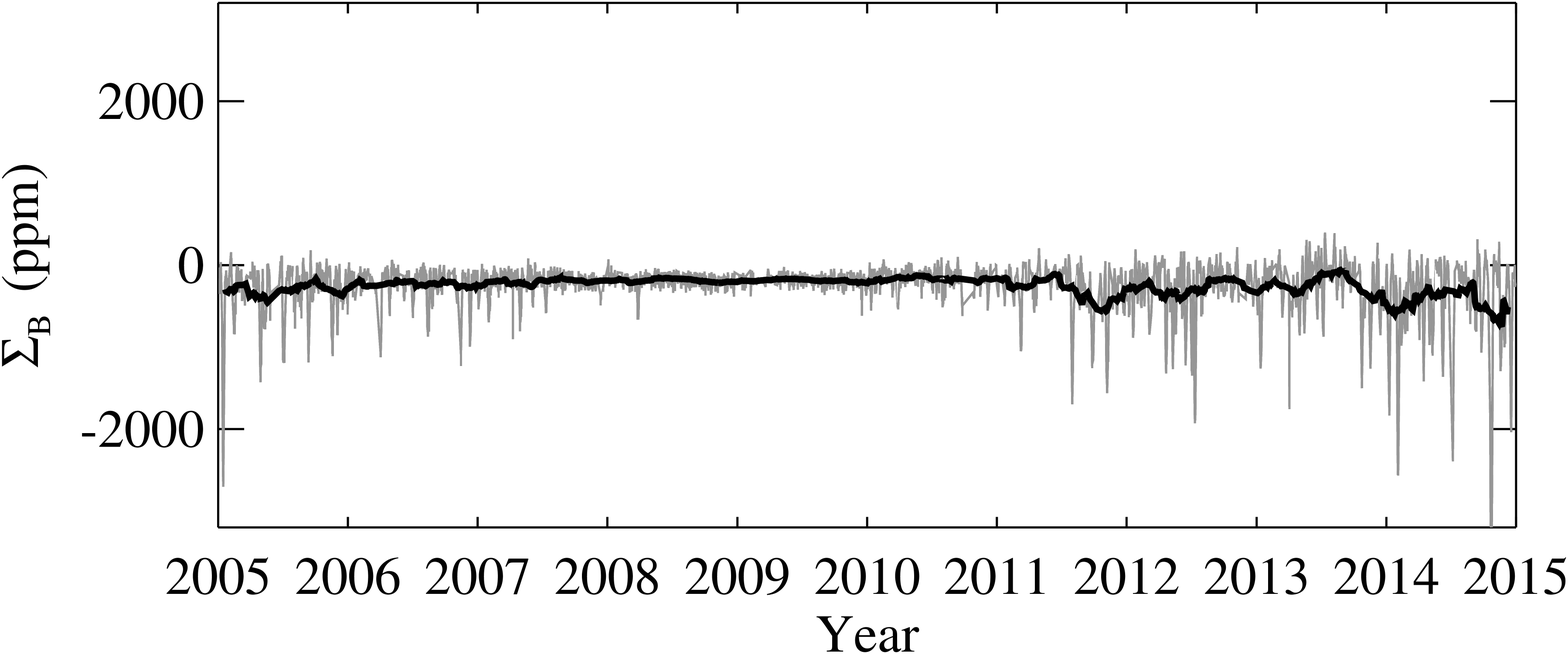}}
	\subfloat{\includegraphics[width=85mm,height=32mm, trim=25mm 22mm 0mm 35mm, clip=true]{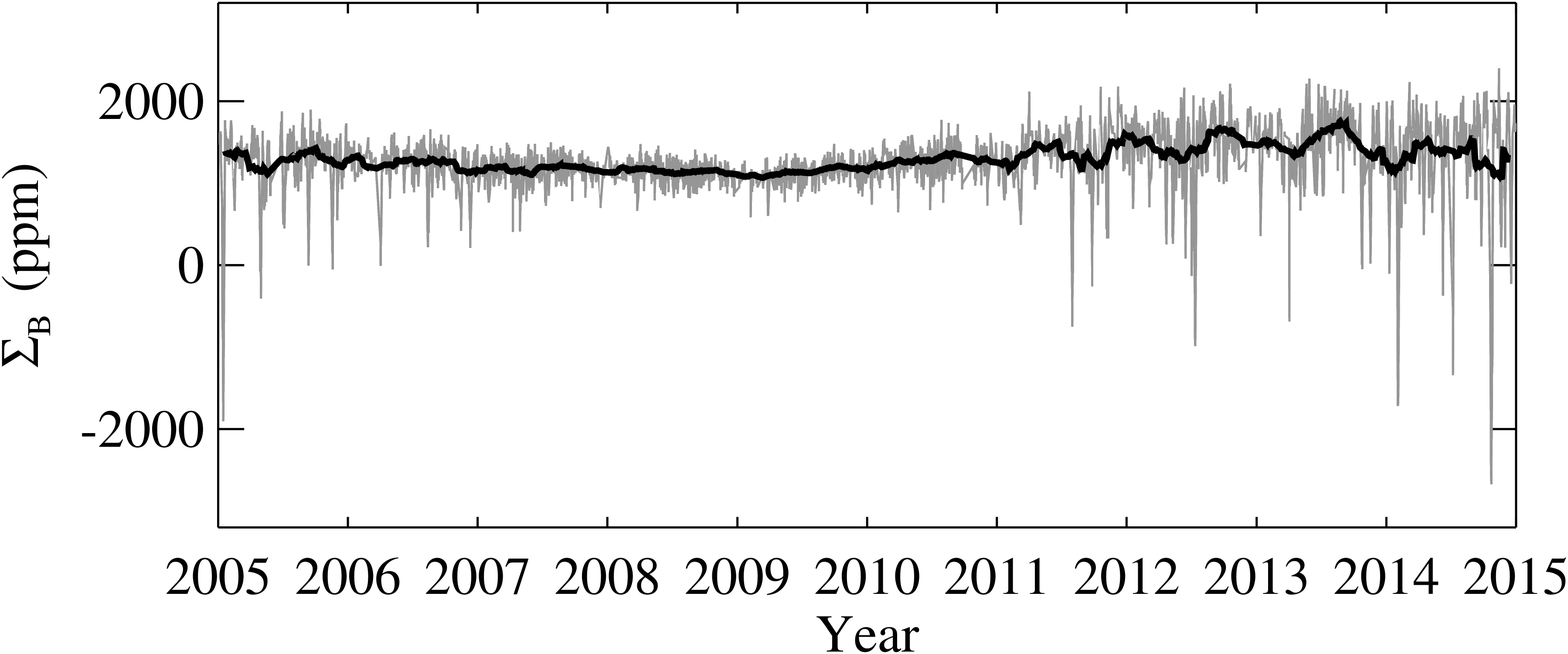}}
\caption{Left: Photometric sum in red (top), $\Sigma_{\rm R}$, and blue (bottom), $\Sigma_{\rm B}$, using the median intensity as a proxy for quiet-sun pixels in the CLV determination. Right: Photometric sums using the median intensity of internetwork pixels only in the CLV.  Bold lines are 81 day running averages to highlight longer-term trends. Note the trend reversal between the two processing schemes.}
\label{total trends}
\end{figure*}

\section{Photometric trends using PSPT data}

The Precision Solar Photometric Telescope (PSPT) at the Mauna Loa Solar Observatory (MLSO) produces daily seeing-limited full-disk images at blue continuum (409.4nm, FWHM 0.3nm), red continuum (607.1nm, FWHM 0.5nm), and CaII K (393.4nm, FWHM 0.3nm) wavelengths\footnote{filter profiles at $\rm{http://lasp.colorado.edu/pspt\_\,access/}$} with up to 0.1\% pixel-to-pixel relative photometric precision~\citep{rast}. We use the red and blue images to measure the solar photospheric continuum contrasts and the CaII K images for magnetic structure identification~\citep{fontenla}, with sunspot umbra and penumbra identified by their red continuum deficit. 

We employ two standard definitions of the quiet-sun to compute the CLV, against which the contrast of each image pixel is measured. The first method employs the median intensity of all pixels in equal area concentric annuli as a somewhat activity insensitive proxy for the quiet-sun intensity. The details of this method are adopted from the work of \citet{walton} and \citet{prem2}. 
The second uses the median intensity of internetwork pixels as identified by the Solar Radiation Physical Modeling (SRPM) semi-empirical thresholding scheme~\citep[and references therein]{fontenla} in aligned CaII K images.  Both schemes employ 100 equal-area annuli between $\mu=0.2$ and 1, where $\mu$ is the cosine of the heliocentric angle.  We restrict our analysis to $\mu$ greater than 0.2, corresponding to $0.975R_{\rm Sun}$, to lesson limb effects where structure identification becomes less robust and the CLV value becomes unreliable due to over enhancement of very near-limb structures. 

Subtracting the CLV profile from the image and normalizing by the disk integrated CLV creates a photometric map, identifying the relative contribution of each pixel to the disk integrated contrast. Summing over all image pixels yields the photometric sum defined by Preminger et al. 2011 as
\begin{equation}
\Sigma = \frac{\sum_{i}\left( I_i - I(\mu_i)\right) }{\sum_{i} I(\mu_i)}\ ,
\end{equation}
where I$_{i}$ is the observed intensity, $\mu_i$ is the cosine of the heliocentric angle, and I($\mu_i$) is the CLV value determined using  one of the two methods described above at the $i$th pixel location. 

Figure 1 plots the photometric sums over the red and blue continuum images, $\Sigma_{\rm R}$ and $\Sigma_{\rm B}$, as a function of time for each of the CLV definitions. Short-term variations result from solar features, primarily sunspots and faculae, crossing the solar disk and changing the solar irradiance on time scales of a few days or weeks. Over-plotted are 81 day running mean values which highlight longer-term trends. As in previous work~\citep{prem1}, when employing the quiet-sun proxy (median intensity of all pixels in annuli)  in the construction of the CLV profile $\Sigma_{\rm R}$ and $\Sigma_{\rm B}$ are out of phase with solar activity.  However, when SRPM internetwork pixels alone are used in the CLV determination the photometric sums are in phase with solar activity.  Long-term trends in the photometric sum are sensitive to the CLV profile employed. (Note: We will address the low amplitude annual variations seen in Figure~1 in Section~3)

Using the SRPM image mask, we decompose the total photometric sum of Figure~1 into the contributions from magnetic structures: internetwork, network, active network, faculae, and sunspots.  
As shown in Figure~2 for the red continuum, the amplitude and phase dependence of sunspots and faculae contrasts show very little sensitivity to the CLV profiles.  Using either profile, disk integrated facular contrasts are in phase with the solar cycle while that of sunspots is out of phase.  The active network contribution is similar to that of faculae, showing an in phase relationship to the solar cycle, but the magnitude of the change with cycle is  significantly larger when the CLV profile is based on quiet-sun pixels. The network contribution is out of phase with the cycle, with trends of approximately the same magnitude, independent of the CLV methodology.  The internetwork dependence on cycle, is not surprisingly flat when the quiet-sun (internetwork) pixels are used to construct the CLV, but shows an out of phase trend with solar cycle when the median proxy is employed. We note that the values presented here are total structure contributions normalized by the full disk area but not normalized to the number of pixels of each structure type, which changes with solar activity.  Thus temporal variations in the photometric sums reflect two possible underlying causes, changes in the fractional area covered by the magnetic structure types and/or changes in the structure contrasts. 

\begin{figure*}[ht!]
	\subfloat{\includegraphics[width=85mm,height=30mm, trim=25mm 45mm 0mm 35mm, clip=true]{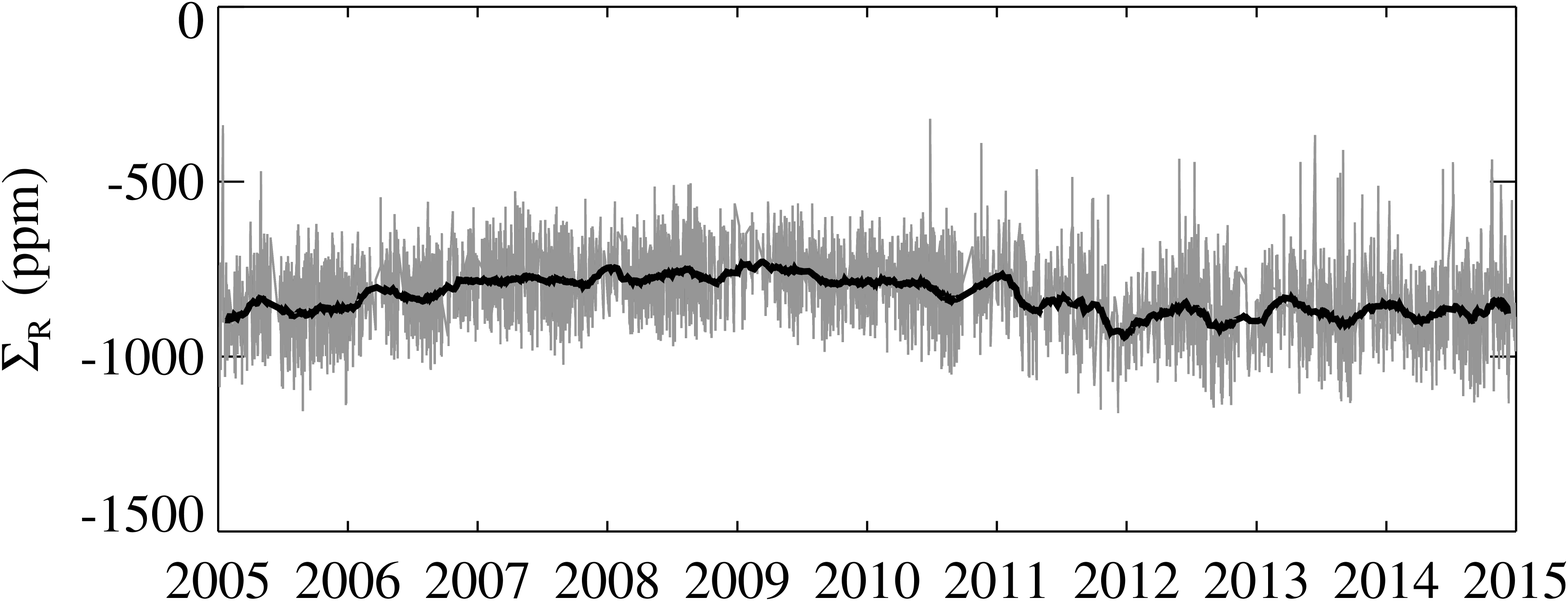}}
	\subfloat{\includegraphics[width=85mm,height=30mm, trim=25mm 45mm 0mm 35mm, clip=true]{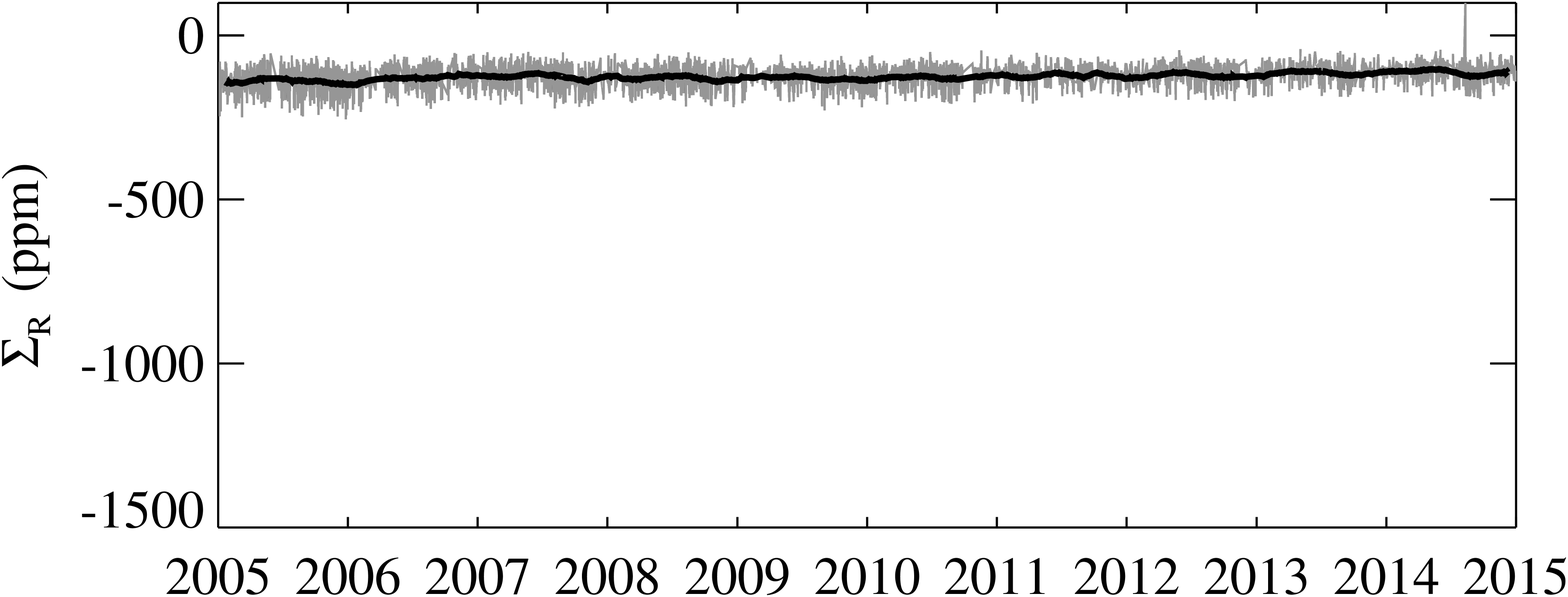}}\\
	\subfloat{\includegraphics[width=85mm,height=30mm, trim=25mm 45mm 0mm 35mm, clip=true]{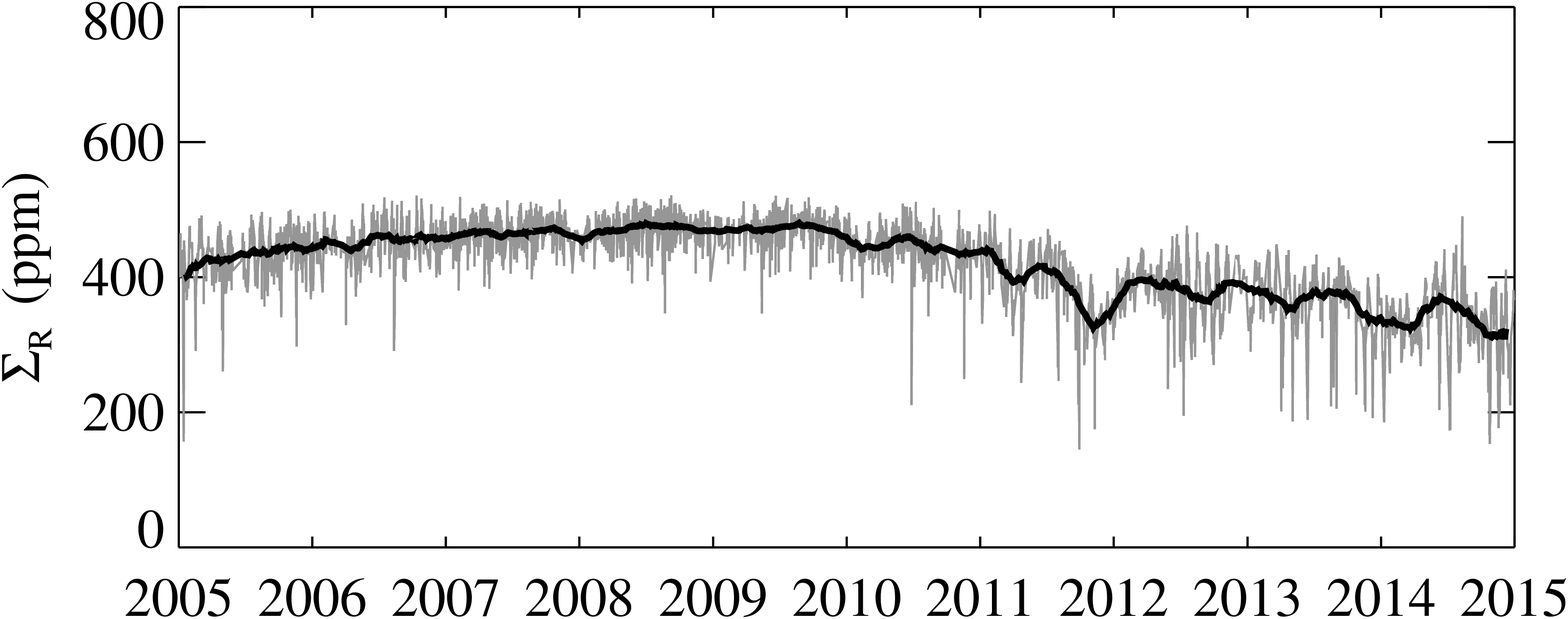}}
	\subfloat{\includegraphics[width=85mm,height=30mm, trim=25mm 45mm 0mm 35mm, clip=true]{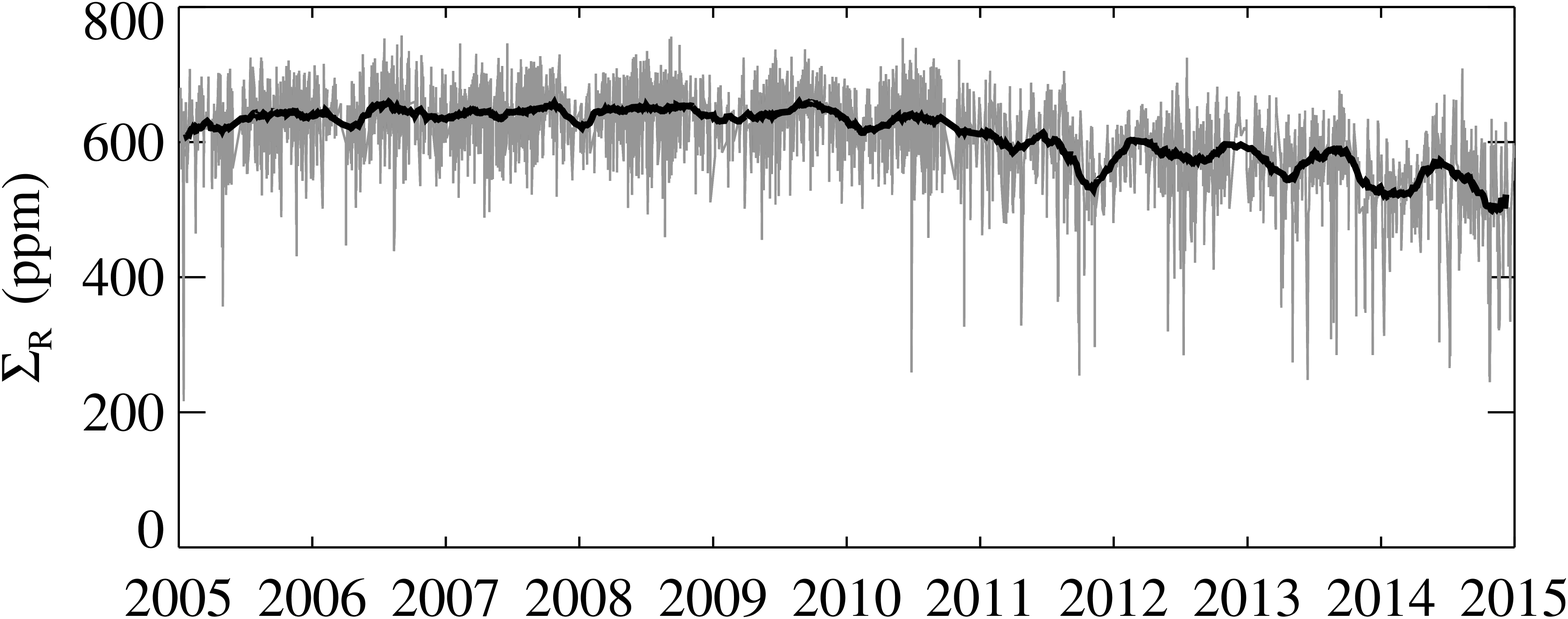}}\\
	\subfloat{\includegraphics[width=85mm,height=30mm, trim=25mm 45mm 0mm 35mm, clip=true]{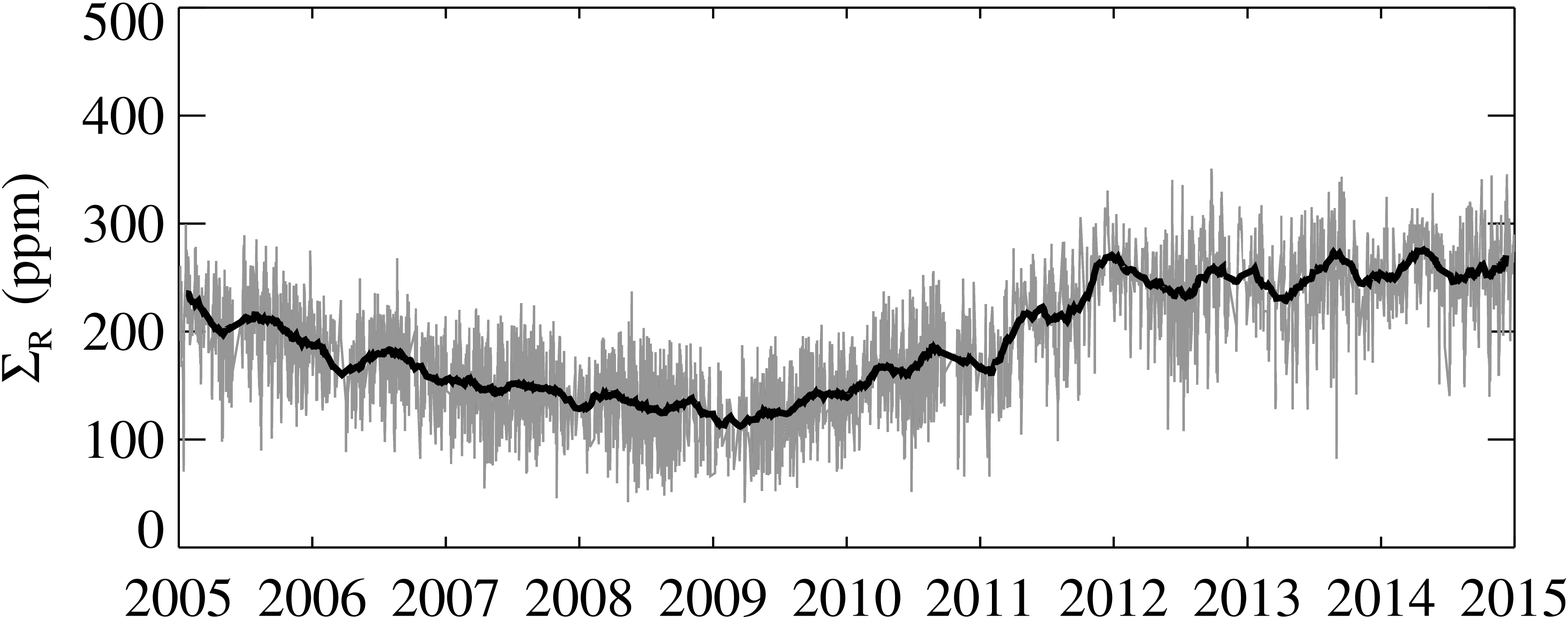}}
	\subfloat{\includegraphics[width=85mm,height=30mm, trim=25mm 45mm 0mm 35mm, clip=true]{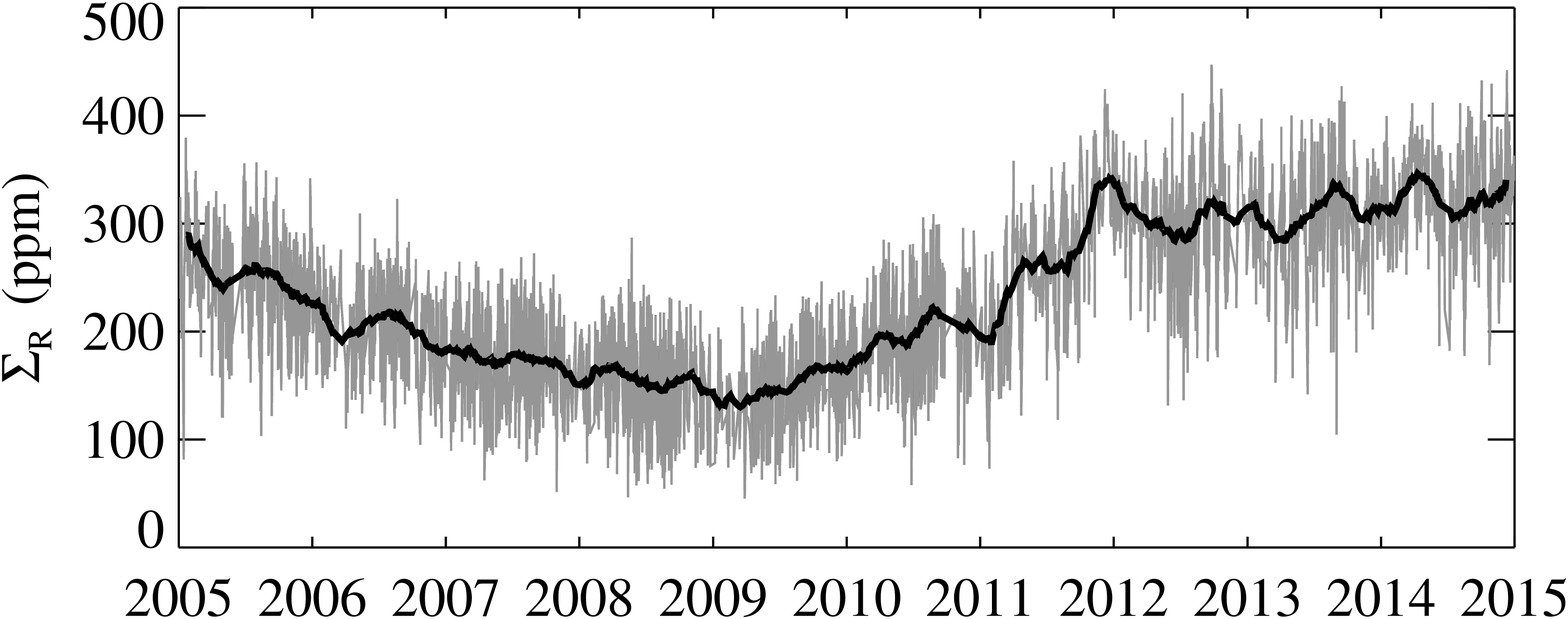}}\\
	\subfloat{\includegraphics[width=85mm,height=30mm, trim=25mm 45mm 0mm 35mm, clip=true]{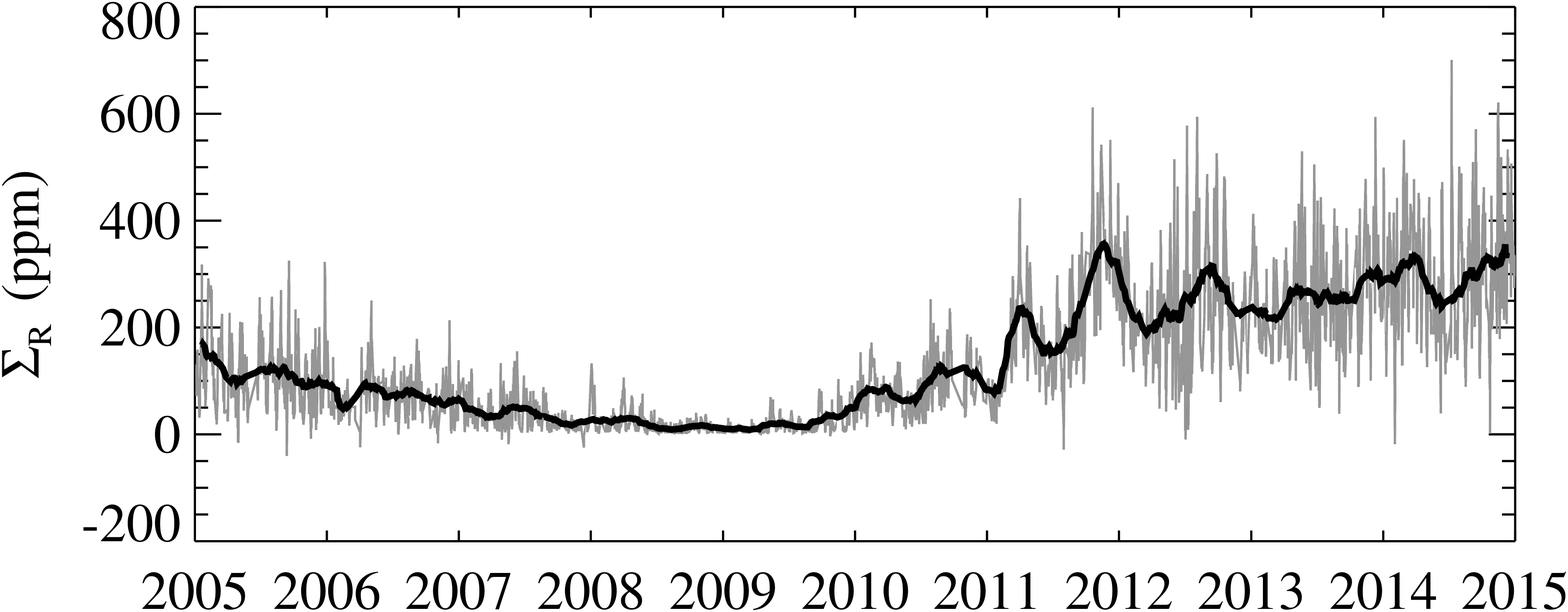}}
	\subfloat{\includegraphics[width=85mm,height=30mm, trim=25mm 45mm 0mm 35mm, clip=true]{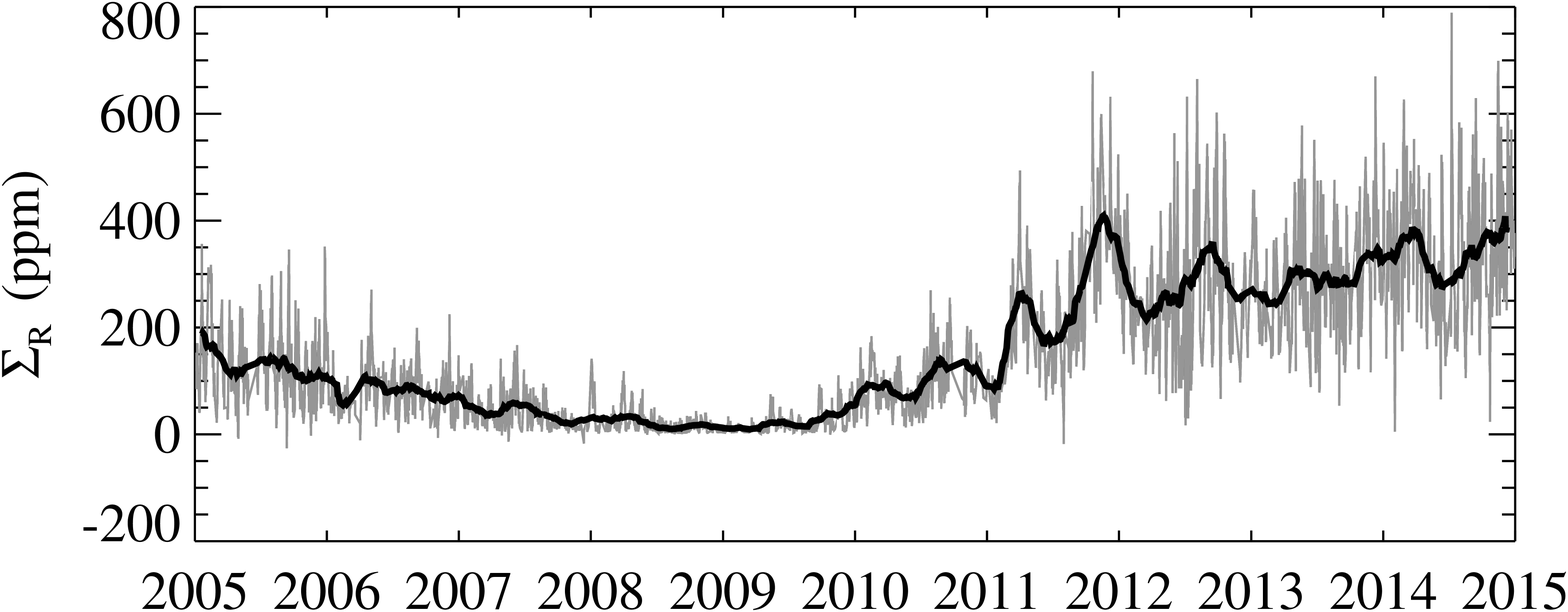}}\\
	\subfloat{\includegraphics[width=85mm,height=32mm, trim=25mm 22mm 0mm 35mm, clip=true]{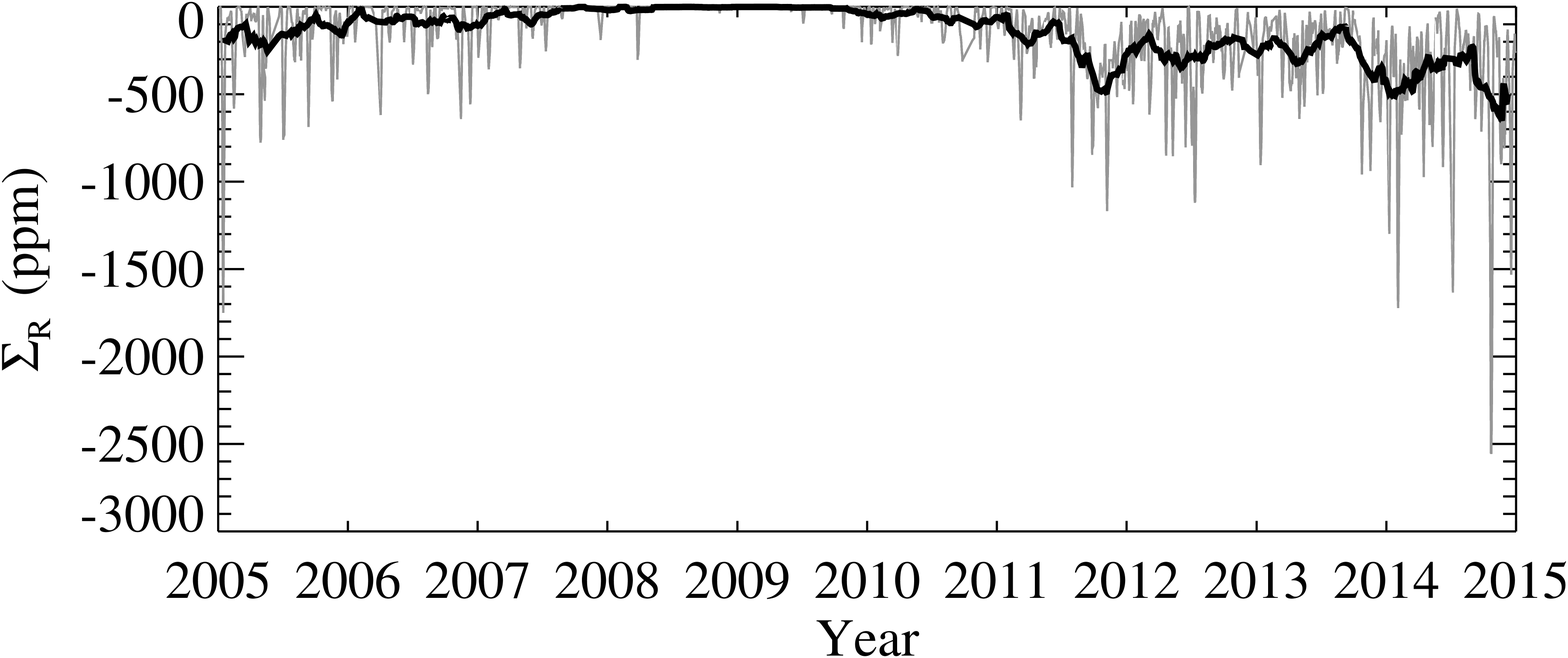}}
	\subfloat{\includegraphics[width=85mm,height=32mm, trim=25mm 22mm 0mm 35mm, clip=true]{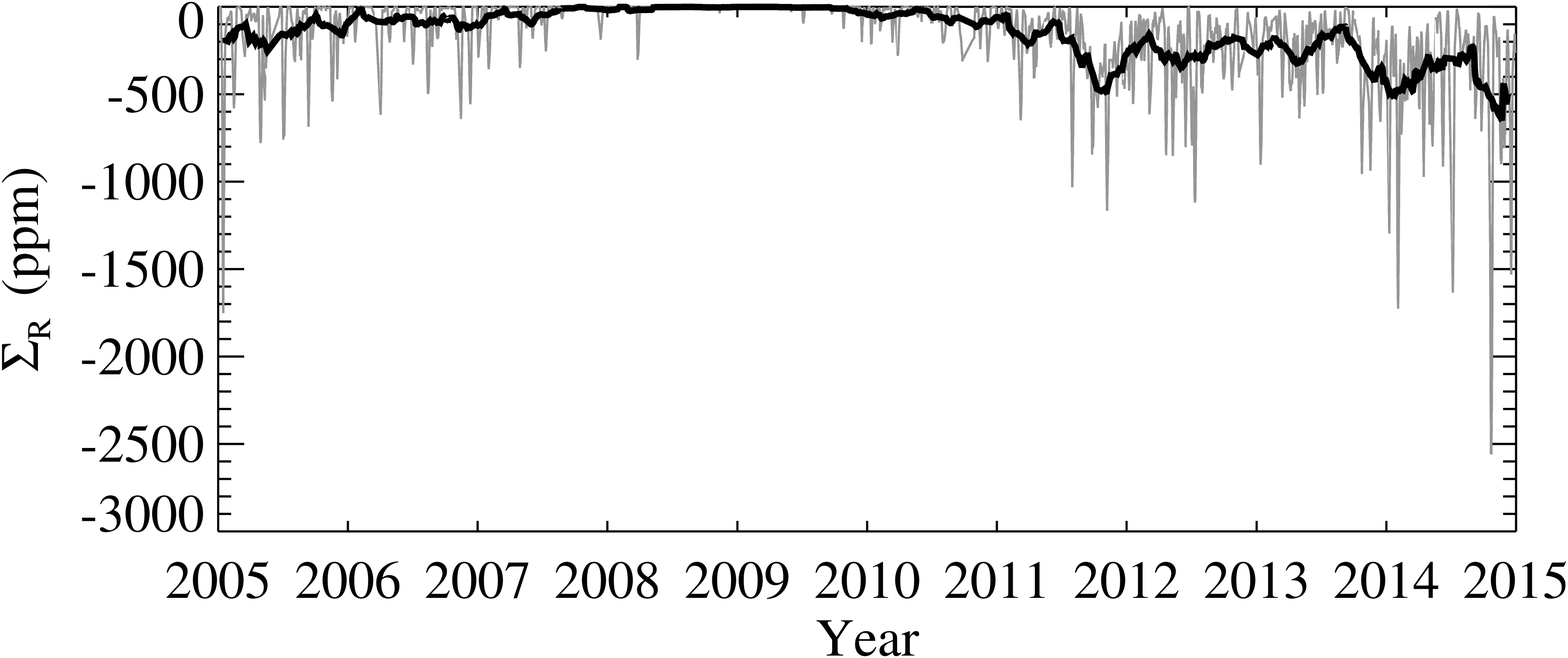}}\\
\caption{Effects of the CLV on photometric trends of SRPM identified features. Left: Photometric sum, $\Sigma_{\rm R}$, of solar elements using the median intensity as a proxy for quiet-sun pixels. Right: Photometric sum using the median intensity of only quiet-sun pixels. From top to bottom: internetwork, network, active network, faculae, and sunspots. Bold lines are 81 day running averages to highlight longer-term trends. Note the trend reversal in the internetwork pixels only between the two processing schemes, which demonstrates the sensitivity of low contrast, high abundance features to the CLV.}
\label{component trends}
\end{figure*}

\section{Contributions of the center-to-limb profile to observed photometric trends}

The differences between the right and left columns of Figures~1 and 2 reflect the sensitivity of the photometric measures to the CLV employed. Since the CLV is constructed using a `quiet-sun' reference and the contrast of each image pixel is measured against the CLV, the quiet-sun reference plays a significant role in the observed contrast trends. The quiet-sun reference can vary over the solar cycle due to changes in the quiet-sun identification, the filling factor of sub-pixel unresolved magnetic structures, or the underlying thermodynamic structure of the atmosphere. 

While we can not differentiate between these underlying causes, we can evaluate how the CLV definitions of the previous section lead to the observed trends. The difference between the total photometric sum obtained when using the CLV constructed from the quiet-sun median value proxy,
$\Sigma_{\rm proxy}$, and the total photometric sum obtained when using the CLV constructed from internetwork pixels, $\Sigma_{\rm quiet-sun}$, can be written as
\begin{equation}
\Sigma_{\rm quiet-sun} - \Sigma_{\rm proxy} = -\delta(\Sigma_{\rm proxy}+1)\ ,
\end{equation}
where 
\begin{equation}
\delta =\frac{\sum\nolimits_{i}\left( I_{\rm quiet-sun}(\mu_i)-I_{\rm proxy}(\mu_i)\right) }{\sum\nolimits_{i}I_{\rm quiet-sun}(\mu_i)}\ 
\end{equation}
is the disk integrated difference in the CLVs applied. The difference in the total photometric sums thus depends linearly on the differences between the CLVs. This implies that the trend reversal in Figure~1 resulted from a cycle dependent difference between the two CLVs of the same order of magnitude as the photometric trends themselves.

The disk-integrated relative difference between the CLVs (equation~3) which gives rise to the trend reversal of Figure~1 includes both the cycle dependent differences and a constant offset due to the mean intensity of the pixels used in the quiet-sun reference.  We are interested in cycle dependent differences between the CLVs and so define $\delta_{\rm measured}$ as $\delta$ minus its temporal average, and plot that disk-integrated quantity in Figure~3.

The disk integrated CLV differences are out of phase with solar cycle and of the same magnitude of the total photometric trends of Figure~1. Since the only difference between the two CLV profiles is the number of resolved magnetic structures included in the quiet-sun reference, the photometric sum deduced from one or both of the quiet-sun references contains measurable cycle-dependent contamination which manifest in the photometric sum via the CLV. 

\begin{figure}[h!]
\centering
	\subfloat{\includegraphics[width=80mm,height=32mm, trim=30mm 20mm 0mm 35mm, clip=true]{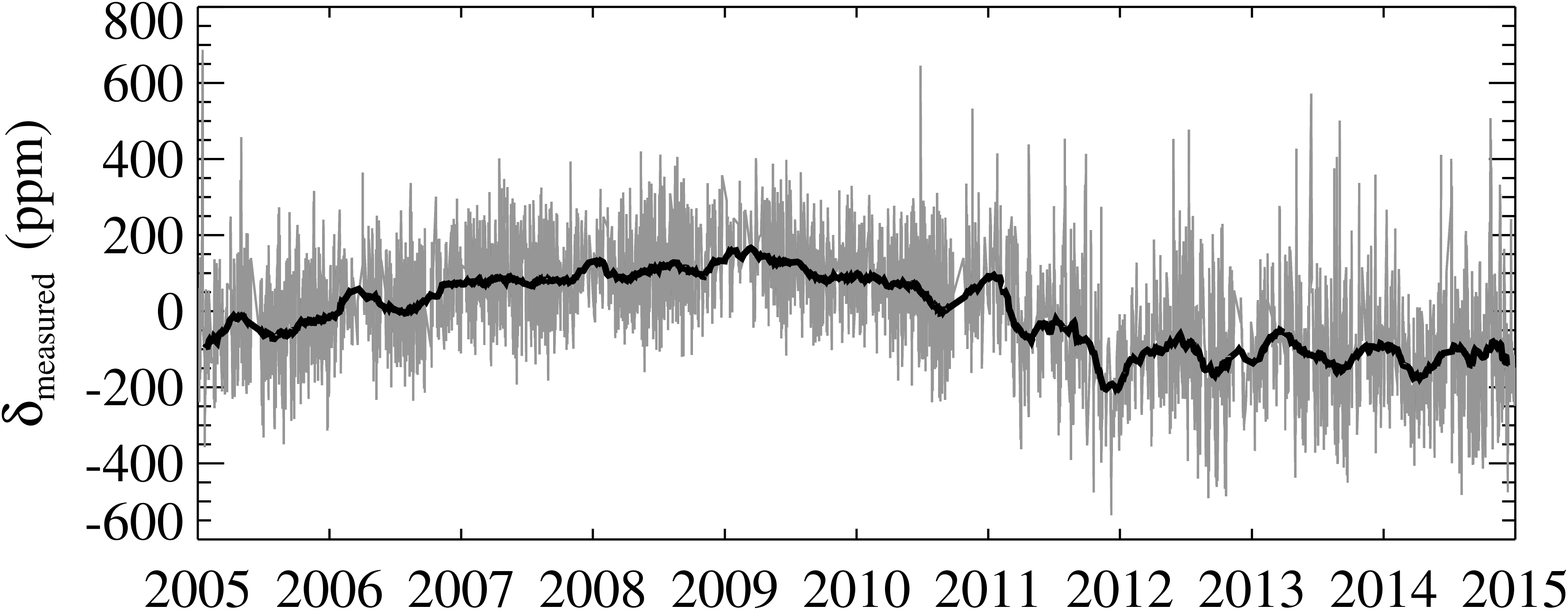}}\\
	\subfloat{\includegraphics[width=80mm,height=32mm, trim=30mm 20mm 0mm 35mm, clip=true]{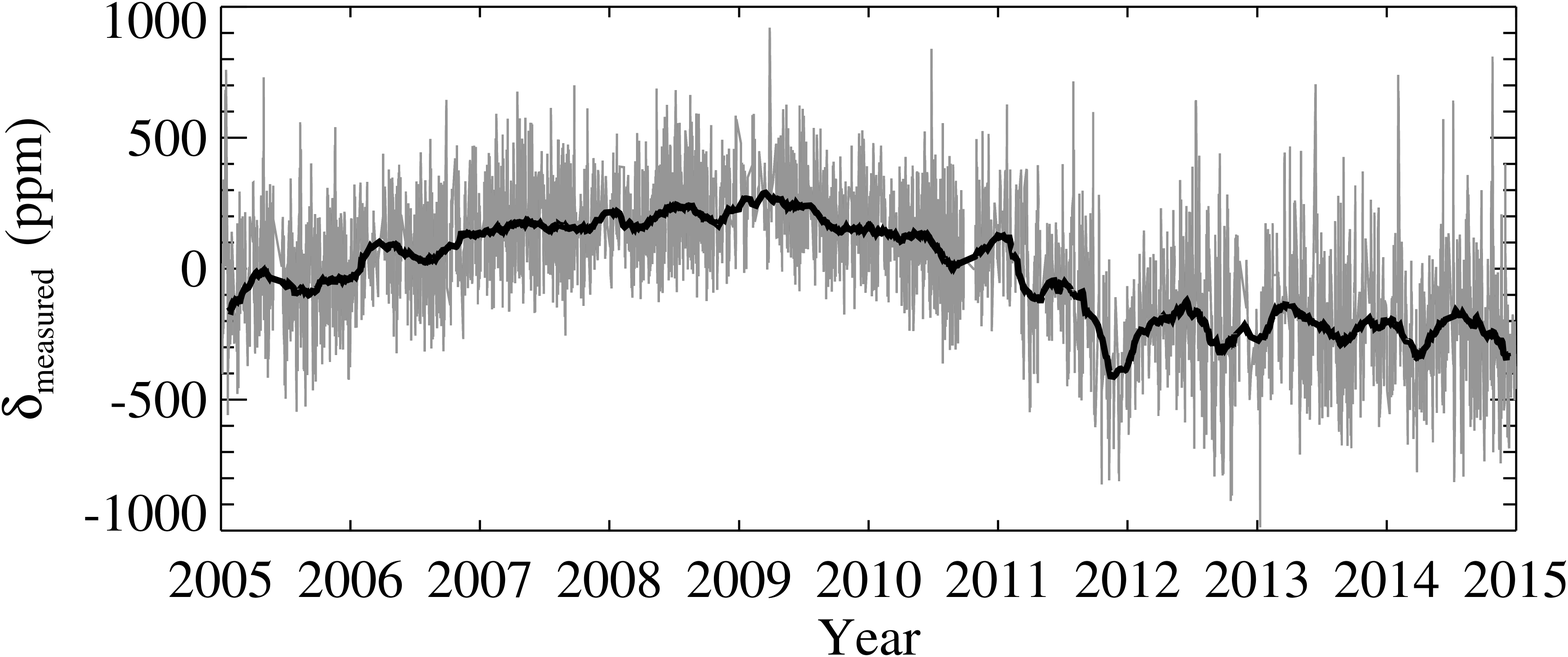}}
\caption{Relative difference between the disk integrated quiet-sun CLV and quiet-sun proxy CLV minus the temporal average, $\delta_{\rm measured}$, in red (top) and blue (bottom). Bold lines are 81 day running averages to highlight longer-term trends. Note the cycle dependence of the relative difference which changes in each annulus.}
\label{CLV difference}
\end{figure}

Evaluating the disk normalized changes between the two CLVs (quiet-sun minus proxy) as a function of disk position illustrates how the underlying quiet-sun references change the CLV shape with solar cycle (Figure~4).  Only the 81-day running averages are plotted to emphasize the longer-term trends. The cycle dependency of the CLV difference in each annulus implies that one or both of the quiet-sun references introduces contamination by magnetically enhanced structures in the CLV.  The differences near the limb are out of phase with solar cycle, while those near disk center are in phase. This indicates that the CLV profiles are changing shape with respect to each other, and this in turn effects any measurement of the center-to-limb profile of magnetic structure contrast on the disk and the contribution of these structures to the photometric sums in Figure~2.

\begin{figure}[h!]
\centering
	\subfloat{\includegraphics[width=80mm,height=32mm, trim=35mm 20mm 0mm 31mm, clip=true]{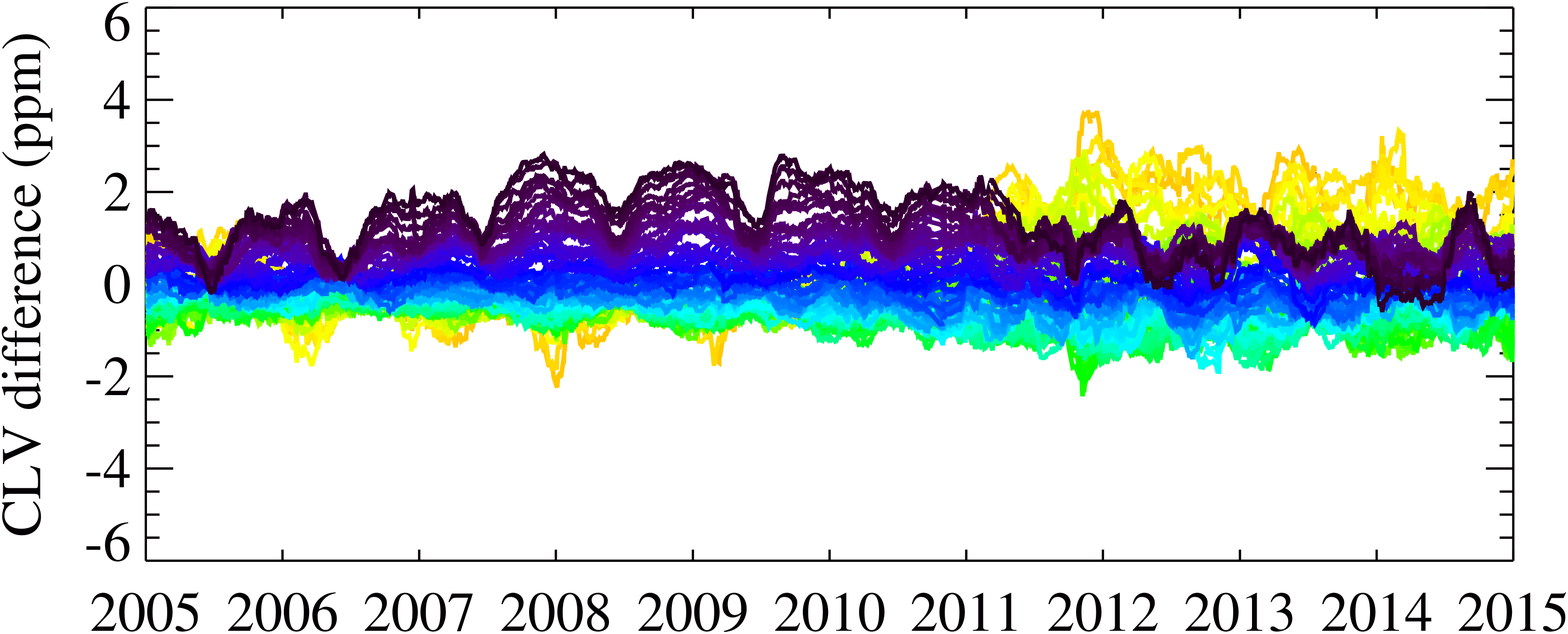}}\\
	\subfloat{\includegraphics[width=80mm,height=32mm, trim=35mm 20mm 0mm 31mm, clip=true]{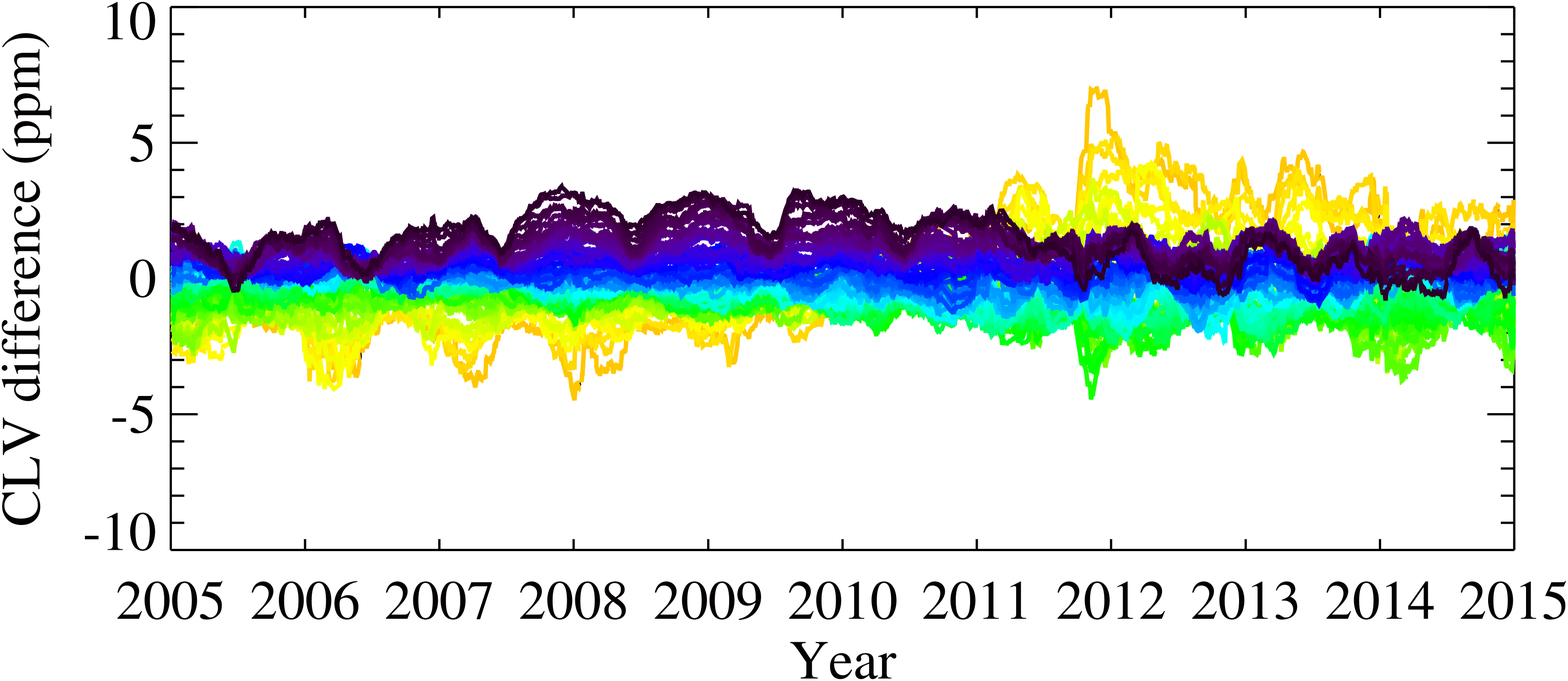}}
\caption{Disk normalized difference between the CLVs in each annulus in red (top) and blue (bottom) from center (yellow) to limb (black) averaged over 81 days to highlight the longer-term trends. Each annulus responds differently to the solar cycle, indicating that the changing shape of the CLV gives rise to the photometric signal changes of Figure~1 and Figure~2.}
\label{CLV mu difference}
\end{figure}

We note the presence of annual variations in Figure~4, and attribute these to changes in the apparent size of the Sun over the Earth's orbital period.  This causes a change in the resolution of the full disk images, with the highest resolution occurring at perihelion, and in turn yields the annual variations in the CLV difference plots.  Varying image resolution causes changes in quiet-sun pixel identification and varying amounts of magnetic element contamination of these.  We have verified this cause by degrading a high quality image by an amount comparable to that cause by the Earth's orbital motion and examining the effect on the CLV profile differences and the photometric sums. We find changes that are consistent with those observed in the time series, though seasonal seeing effects may also contribute.

Using the magnitudes of the long-term trends from Figure~2, we can estimate what cycle dependent variability in the CLV would account for the observed trends in the structure specific photometric sums.  This is a minimum uncertainty threshold for confidence that those trends are real and independent of the CLV algorithm employed. 
From equation~2, the change in the integrated CLV necessary to account for the trends of the magnetic structures is 
\begin{equation}
\delta_{\rm structure} = \frac{1}{n_{\rm structure}}\frac{\Sigma_{\rm max}}{\Sigma_{\rm max}+1}\ ,
\end{equation}
where $n_{\rm structure}$ is the fraction of disk pixels covered by the structure and $\Sigma_{\rm max}$ is the estimated maximum magnitude of the photometric trend for that structure in Figure~2 ignoring the mean offsets. For simplicity, this assumes that error in the CLV is distributed uniformly over disk position. 
Estimated values of $\Sigma_{\rm max}$, $n_{\rm structure}$, and the corresponding $\delta_{\rm structure}$ from equation~4 are shown in Table~1.  The sensitivity of the measured long term trends in the structure specific photometric sums to errors in the CLV profile depends on the structure contrast, with high-abundance low-contrast structure trends being most uncertain.    

\section{Effect of further removing magnetic activity in the quiet-sun reference}

A few hundred ppm change in the integrated CLV is sufficient to account for the longterm variation of the total disk, internetwork, and network photometric sums, illustrating the sensitivity of the low contrast, high abundance full disk, internetwork, and network trends to uncertainty in the CLV profile.  Much larger errors in the CLV are required to account for long term trends of the high contrast active network, faculae, and sunspot components.  As we have seen, the differences between the CLV profiles based on median or SRPM quiet-sun intensities suggests that one or 
both of the quiet-sun proxies used in construction of the CLV profile is likely contaminated by cycle dependent magnetic activity. This motivates further masking of the magnetic activity, below that of the SRPM quiet-sun definition, as a means of reducing cycle dependent variability in the reference CLV. 

\begin{table}[t!]\small\centering
\begin{tabular}{|l |l |l |l | p{2cm} |}
\hline
Structure & $n_{\rm structure}$ & $\Sigma_{\rm max}$ & $\delta_{\rm structure}$ \\ \hline
Total Disk & $1.0$ & 500 ppm & 500 ppm \\ \hline
Internetwork & $0.75$ & 250 ppm & 330 ppm \\ \hline
Network & $0.20$ & 100 ppm & 500 ppm \\ \hline
Active network & $0.04$ & 200 ppm & 5000 ppm \\ \hline
Faculae & $0.01$ & 300 ppm & 30000 ppm \\ \hline
Sunspots & $0.001$ & 500 ppm & 500000 ppm \\ \hline 
\end{tabular}
\caption{Magnetic structure fraction of solar disk pixels, $n_{\rm structure}$, estimated magnitude of photometric trends, $\Sigma_{\rm max}$, and estimations of required error in CLV to remove trends, $\delta_{\rm structure}$. The highest abundance structures are most sensitive to CLV error.}
\end{table}

Using co-aligned red and Ca II K images, we compute the CLV for the red continuum images using the internetwork pixels with the least magnetic flux; we employ co-temporal and co-aligned Ca II K images as a proxy for magnetic flux density~\citep{ortizrast}. We compute ten different CLV profiles as in Section~2, but constructed using the median continuum intensity of the internetwork pixels after applying Ca II K thresholds of increasing severity, including from 90\% to the 1\% of the darkest Ca II K internetwork pixels. 

The resulting 81 day running average photometric sums (minus their temporal mean values) $\Sigma_{\rm trend}$ are plotted in Figure~5. The internetwork and network trends are out of phase with solar cycle independent of the threshold level applied. As the magnetic flux of the quiet-sun reference is decreased by increasing the severity of the Ca II K mask, the trends become more pronounced.  Active network and faculae remain in phase with the solar cycle, and these trends increase with decreasing magnetic flux in the quiet-sun reference. The sunspot trends display little sensitivity to the thresholds employed in the CLV definition, consistent with equation~4 and Table~1 indicating that very large changes in the CLV are necessary to change the trends of sunspots. 

It is important to note that, while the amplitude of the long term trends of the individual magnetic structures are enhanced with increasingly severe magnetic masking in construction of the CLV, the cycle related trend in the {\it total photometric sum} decreases with increasing severity of the mask. This is again due to the greater sensitivity of the low contrast structures, the change of the internetwork and network trends is larger in magnitude than the oppositely signed change in the active network and faculae contributions. 

\begin{figure*}[ht!]
	\subfloat{\includegraphics[width=85mm,height=30mm, trim=25mm 45mm 0mm 35mm, clip=true]{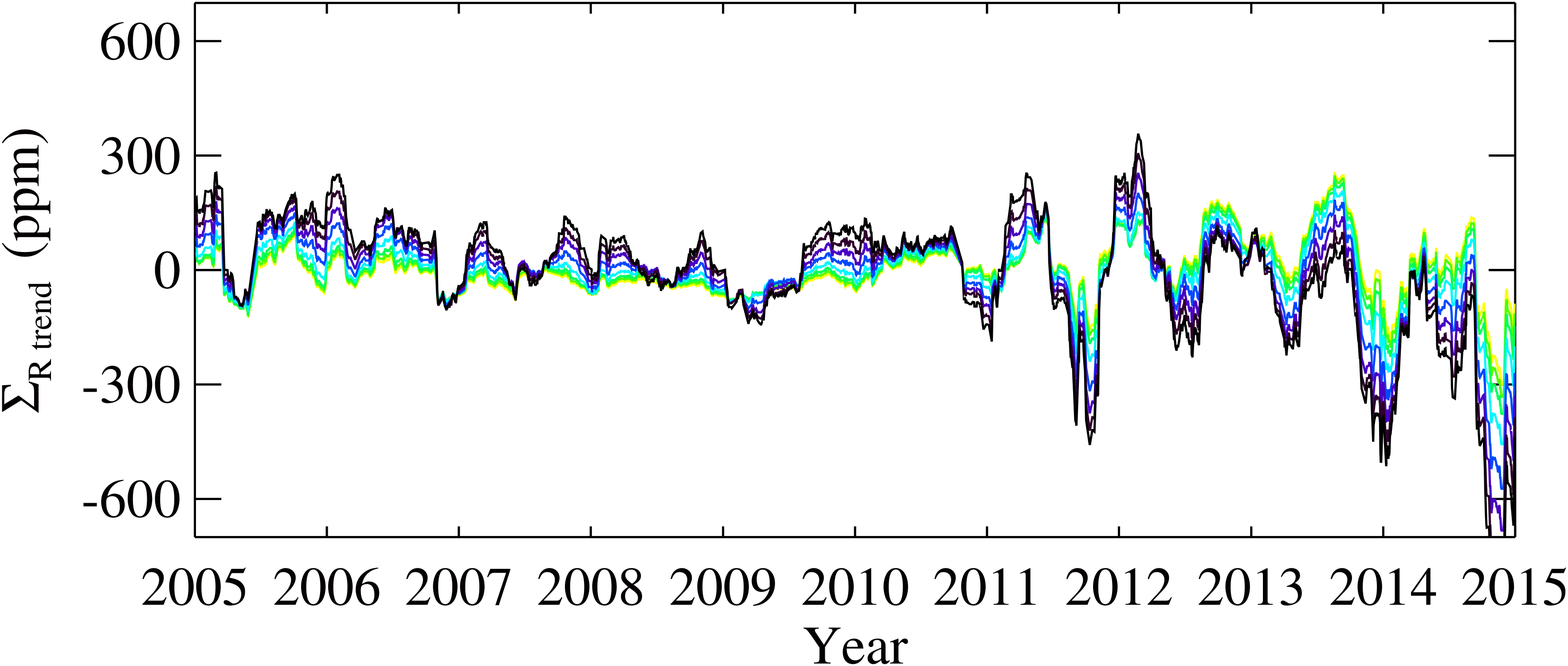}}
	\subfloat{\includegraphics[width=85mm,height=30mm, trim=25mm 45mm 0mm 35mm, clip=true]{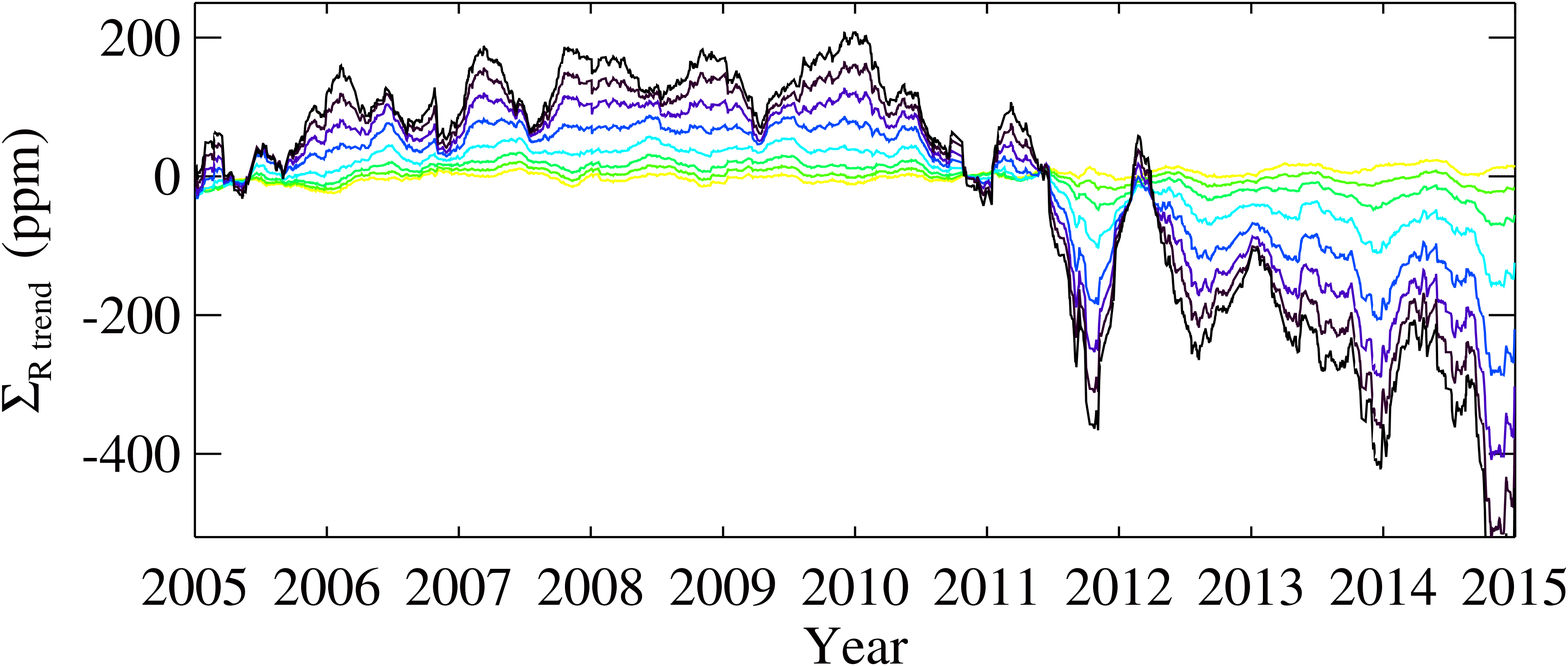}}\\
	\subfloat{\includegraphics[width=85mm,height=30mm, trim=25mm 45mm 0mm 35mm, clip=true]{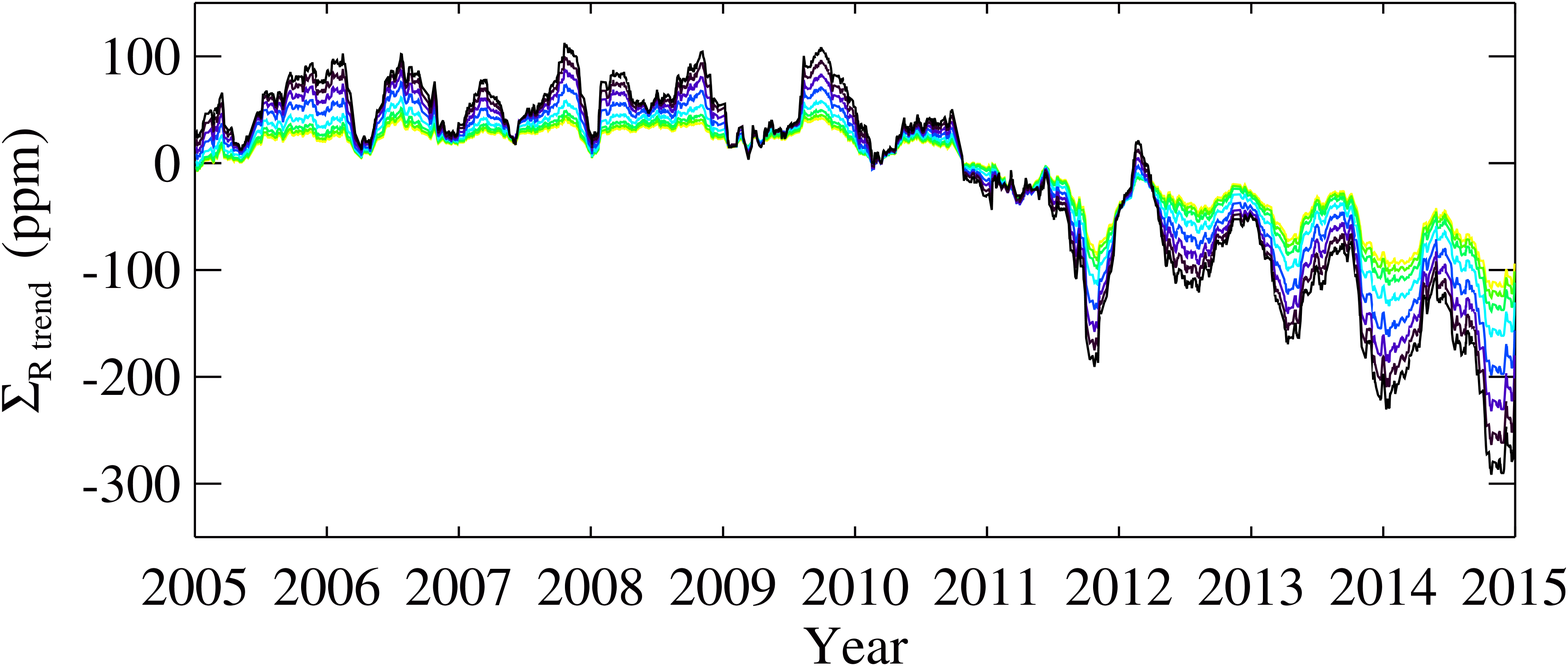}}
	\subfloat{\includegraphics[width=85mm,height=30mm, trim=25mm 45mm 0mm 35mm, clip=true]{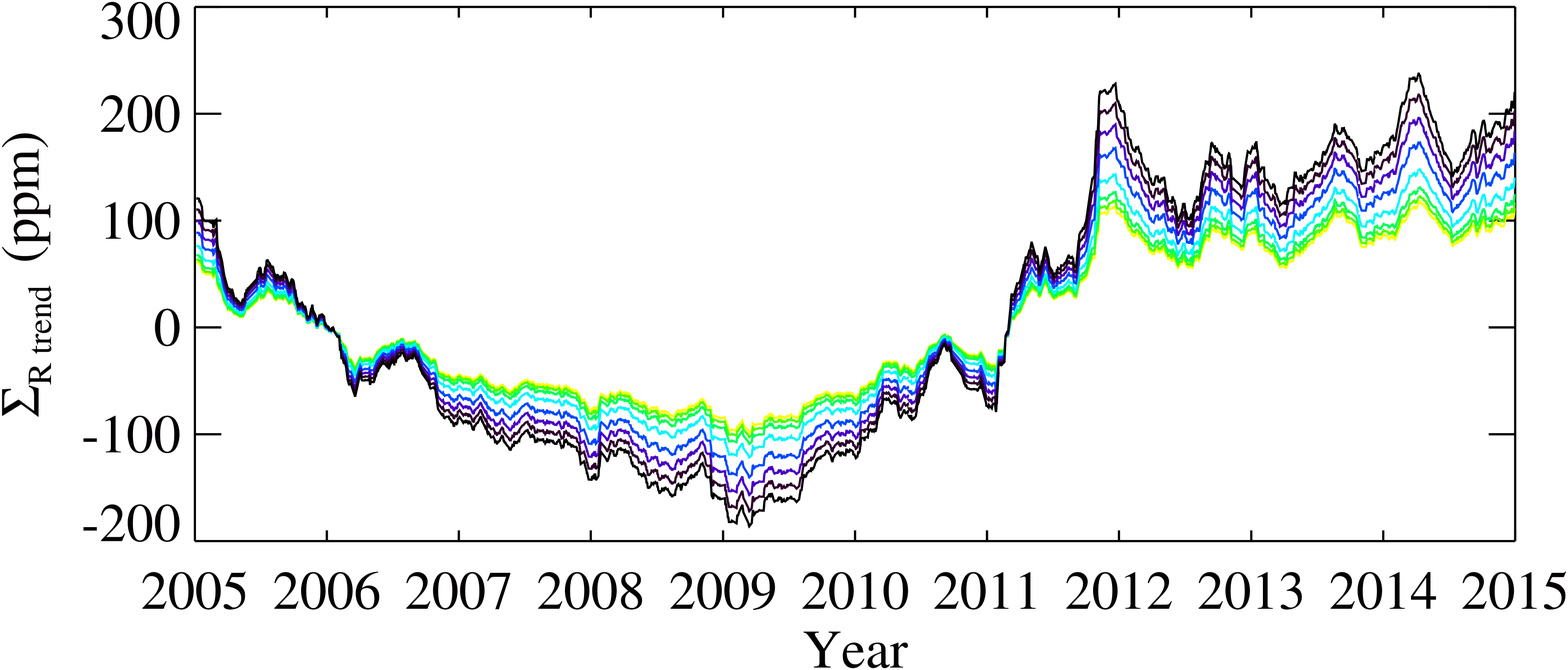}}\\
	\subfloat{\includegraphics[width=85mm,height=30mm, trim=25mm 45mm 0mm 35mm, clip=true]{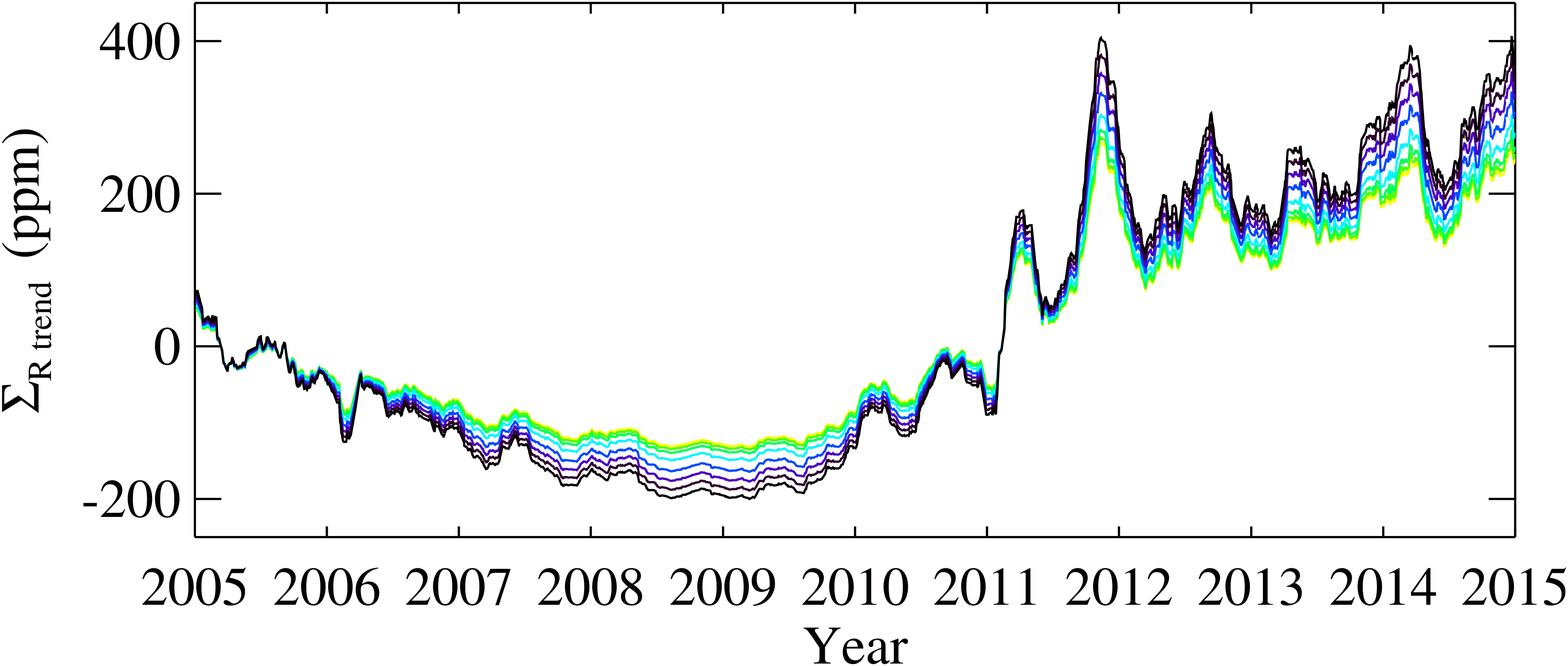}}
	\subfloat{\includegraphics[width=85mm,height=30mm, trim=25mm 45mm 0mm 35mm, clip=true]{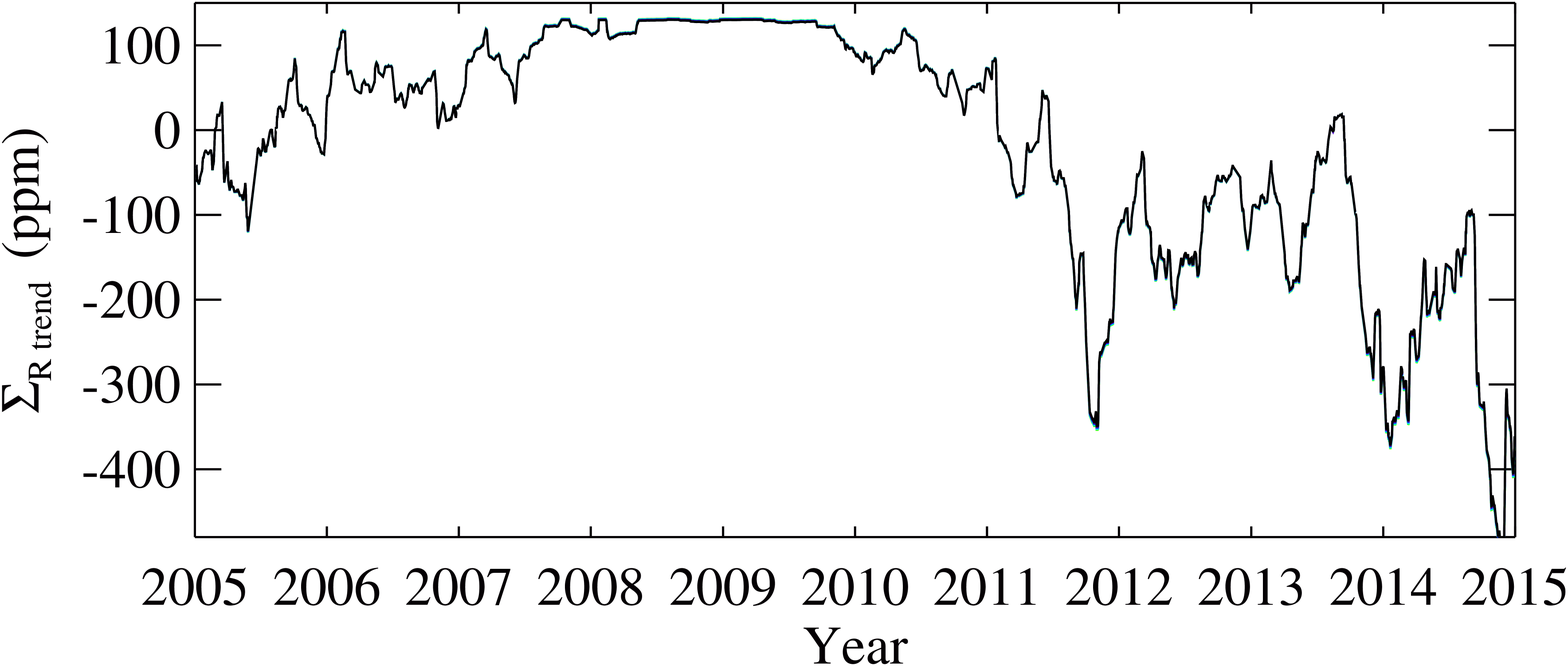}}\\
\caption{Photometric sums minus their temporal averages, $\Sigma_{\rm R trend}$, averaged over 81 days using various levels of magnetic thresholding from Calcium images in the CLV to remove magnetically enhanced pixels from the quiet-sun reference. Trends are shown for the total photometric sum (top left), internetwork (top right), network (middle left), active network (middle right), faculae \& plage (bottom left), and sunspots (bottom right). Note the enhancement of all structure trends as more magnetic flux is masked, while the total photometric trend diminishes.}
\label{threshold trends}
\end{figure*}

As in Section~3, we examine the changes in the continuum CLV shape that leads to the enhanced trends with decreasing internetwork Ca II K intensity defining the reference pixels. This is shown in Figure~6 as the disk normalized difference between the 50\% internetwork  threshold CLV and the 100\% internetwork pixel CLV. Unlike in Figure~4, which compares the difference between the median value and 100\% internetwork CLV results, the  difference between the 50\% threshold CLV and the 100\% internetwork constructed CLV shows little solar cycle dependence. Instead it shows large offsets, increasing with decreasing $\mu$ (toward limb) but nearly constant in time (aside from annual variability discussed below). This behavior is observed for all threshold levels, with the magnitude of the offsets increasing with decreasing magnetic flux levels. The center-to-limb profile becomes increasingly more limb darkened as the masking becomes more severe; the darkest internetwork pixels are more limb darkened than the median internetwork. 

The increasingly limb darkened CLVs when applied to the image data yield the structure trends of Figure~5. Faculae are increasingly limb brightened, while the network and internetwork, inherently of more uniform intensity across the disk, have enhanced contrast at the limb but slightly lower contrast near at the center relative to the CLV profile.  This yields the enhanced out of phase trends in the photometric sum. 
The magnitudes of photometric sum trends, excluding that of sunspots, are highly dependent on the center-to-limb profile of the somewhat arbitrary quiet-sun reference. The true center-to-limb profiles of magnetic structures, their contributions to solar cycle irradiance trends, and the role of the quiet-sun itself can only be determined via absolute imaging radiometry.  

\begin{figure}[h!]
	\subfloat{\includegraphics[width=80mm,height=32mm,trim=35mm 20mm 0mm 30mm, clip=true]{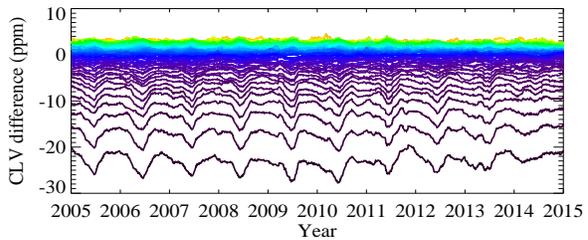}}
\caption{Disk normalized difference between the 50\% internetwork magnetic threshold and the internetwork CLV in each annulus from center (yellow) to limb (black) averaged over 81 days to highlight the longer-term trends. Note how the annuli differences have little dependency on solar cycle and the 50\% threshold CLV is strongly limb-darkened compared to the internetwork CLV. }
\label{int50}
\end{figure}

We again note the annual variations in Figure~6.  These have the same cause as that described in Section~3. Image blurring acts to reduce limb darkening by contaminating the CLV with magnetically bright structures on the limb. This blurring is most prominent during apoapsis and results in brightening of the CLV from neighboring pixels, and more so for the more severely thresholded CLVs which are more sensitive to contamination from blurring. The enhancement of the annual variations in the time series of low contrast structures in Figure~5 demonstrates that the CLV sensitivity to image blurring most effects the low contrast structures, which are most sensitive to changes in the CLV.

\section{Conclusion}

We examined the influence of the CLV profile on contrast trends deduced from red and blue continuum PSPT images, and found that the observed cycle long trends are sensitive to underlying assumptions in the construction of the center-to-limb profile against which the contrast is measured. Using the method of \citet{prem1}, taking the median value of all pixels on the disk as a proxy for the quiet-sun intensity in the construction of the CLV, we obtain results that agree with those published; the integrated continuum contrast is out of phase with solar activity. However, using a CLV which fits the SRPM identified quiet-sun pixels yields the opposite result, an in phase variation with cycle. Moreover, the amplitude of the signal decreases with increasingly severe masking in the quiet sun definition.  Thus, the integrated contrast (photometric sum) is highly sensitive to the CLV profile employed, and variations in the integrated contrast with solar cycle depend on the CLV definition and its consequent variation in time.   Measurements of temporal variation in the photometric sum can not contribute to the resolution of the discrepancy between VIRGO and SIM assessments of the phase of solar spectral irradiance variations as there is no unique or robust way to define the CLV profile against which the photometric sum measurements are made.  

The photometric contributions of specific magnetic structures also vary with the CLV profile employed.  Internetwork and network contributions are most sensitive to variations in the CLV due to their low contrast and high abundance, and this sensitivity underlies the opposing photometric trends obtained with the two processing schemes.  The photometric contribution of  active network, faculae, and sunspots are less sensitive to the underlying CLV profile used. The contrast trends of the dimmest magnetic features are thus the least well determined, yet, as they are the most abundant, they are key to understanding long term irradiance variability. 

Differences between two CLVs results in linear differences in the photometric trends. Cycle dependent changes in the quiet-sun reference from which the CLVs are constructed introduced cycle dependent differences in the CLVs of the same magnitude as the photometric trends, which resulted in the observed trend contradictions. Additionally, we found that the CLVs changed shape relative to each other over the solar cycle suggesting that the quiet-sun references upon which the CLV profiles are based are likely contaminated by magnetic structures in the reference pixels. Employing a quiet-sun reference that reduces the contribution of magnetic flux by increasingly severe masking based on Ca II K intensity reduces solar cycle dependent magnetic contamination of the center-to-limb profile and changes its shape in a way that yields a) an enhancement of individual structure trends and b) a cancelation between them that diminishes the disk integrated contrast (photometric sum) trend.

Analyzing the veracity of a selected quiet-sun reference and its corresponding CLV is non-trivial. The CLV can change with cycle because of underlying contributions from resolved or unresolved magnetic structures;  it includes the center-to-limb variation of structures included intentionally or unintentionally in the quiet-sun reference definition. As manifested by the annual variations apparent in Figure~4 and Figure~6, small changes in image resolution can result in significant changes in the CLV. These changes are caused by consequent quiet-sun reference identification errors and continuum intensity values that include at lower resolution greater contribution from unresolved magnetic flux elements.
While we demonstrated that the contrast trends found using severe thresholding in the CLV construction are less contaminated by cycle dependent magnetic flux variations, we caution that the observed trends can still not distinguish underlying physical causes, most importantly the possible role of any true quiet-sun in spectral irradiance variations, particularly long term variations.

The contrast trends of the internetwork, network, active network, and the full disk sums are most susceptible to changes in the definition of the CLV.  Determining whether cycle related changes are due to changes in the radiative output or filling factor of these components cannot be determined using relative photometric measurements alone, since all such measures are dependent on a arbitrary definition of a quiet-sun reference component.  We stress the need for space-based absolute spectral irradiance imaging to determine the radiative contributions of low contrast components such as internetwork and network to solar spectral irradiance variability. Observations of this kind will enable studies of the radiative properties of magnetic structures and their surroundings as a function of disk position and thus viewing angle without contamination from an uncertain CLV.   

\acknowledgements
This material is based upon work supported by the National Science Foundation Graduate Research Fellowship Program under Grant No. DGE 1144083 and by NASA award number NNX12AB35G.

\clearpage

\clearpage


\begin{thebibliography}{}
\bibitem[Ball et al.(2012)]{ball} Ball, T. W., Unruh, Y. C., Krivova, N. A., et al. 2012, \aap, 541, 27
\bibitem[Ermolli et al.(2003)]{ermolli} Ermolli, I., Berrilli, F., \& Florio, A.  \aap, 412, 857–864, 2003
\bibitem[Fontenla et al.(2009)]{fontenla} Fontenla, J. M., Curdt, W., Haberreitter, M., Harder, J., \& Tian, H., 2009, ApJ, 707, 482-502
\bibitem[Fontenla et al.(2011)]{fon11} Fontenla, J.M., Harder, J., Livingston, W., Snow, M., \& Woods, T., 2011, JGR, 116, D20108 
\bibitem[Foukal et al.(2004)]{foukal} Foukal, P., Bernasconi, P., Eaton, H., Rust, D. 2004, ApJ, 611:L57-L60
\bibitem[Frohlich(2009)]{frohlich} Frohlich, C. 2009, \aap, 501, L27-L30
\bibitem[Frohlich \& Lean(2004)]{frohlichlean} Frohlich, C. \& Lean, J. 2004, \aap, 12, 273-320
\bibitem[Harder et al.(2009)]{harder} Harder, J. W., Fontenla, J. M., Pilewski, P., Richard, E. C., \& Woods, T. N. 2009, GeoRL, 36, L07801
\bibitem[Krivova et al.(2006)]{krivova2006} Krivova, N. A., Solanki, S. K., \& Floyd, L., 2006, \aap, 452, 631-639
\bibitem[Kuhn et al.(1998)]{kuhn1998} Kuhn, J. R., Bush, R., Scherrer, P., \& Scheick, X., 1998, Nature, 392, 155
\bibitem[Lean(2000)]{lean} Lean, J. L. 2000, GeoRL, 27, 2425-2428
\bibitem[Ortiz(2005)]{ortiz} Ortiz, A., 2005, AdSpR, 35, 350
\bibitem[Ortiz \& Rast(2005)]{ortizrast} Ortiz, A. \& Rast, M. 2005, Mem SAIt. Vol. 76, 1018
\bibitem[Preminger et al.(2002)]{prem2} Preminger, D. G., Walton, S. R., \& Chapman, G. A. 2002, JGRA, 107, 1354
\bibitem[Preminger et al.(2011)]{prem1} Preminger, D. G., Chapman, G. A., \& Cookson, A. M. 2011, ApJ, 739, L45
\bibitem[Rast et al.(2008)]{rast} Rast, M. P., Ortiz, A., \& Meisner, R. W. 2008, ApJ, 673 1209-1217
\bibitem[Thaler \& Spruit(2014)]{thaler} Thaler, I. \& Spruit, H. C. 2014 \aap, 566 A11
\bibitem[Walton et al.(1998)]{walton} Walton, S. R., Chapman, G. A., Cookson, A. M., Dobias, J. J., \& Preminger, D. G. 1998, SoPh,  179, 31
\bibitem[Wehrli et al.(2013)]{wehrli} Wehrli, C., Schmutz, W., \& Shapiro, A. I. 2013, \aap, 556L, 3



\end{thebibliography}
\end{document}